\documentclass[fleqn,usenatbib]{mnras}

\usepackage{newtxtext,newtxmath}
\usepackage[T1]{fontenc}
\usepackage{ulem}
\DeclareRobustCommand{\VAN}[3]{#2}
\let\VANthebibliography\thebibliography
\def\thebibliography{\DeclareRobustCommand{\VAN}[3]{##3}\VANthebibliography}

\usepackage{graphicx}	% Including figure files
\usepackage{amsmath}	% Advanced maths commands
\title[AGN Classification]{Fermi LAT AGN classification using supervised machine learning}

\author[Cooper et al. 2022]{
Nathaniel Cooper,$^{1}$\thanks{E-mail: n.j.cooper137@gmail.com}
Maria Giovanna Dainotti,$^{2,3,4}$
Aditya Narendra,$^{5,6}$
Ioannis Liodakis$^{7}$,
Malgorzata Bogdan$^{8,9}$
\\
$^{1}$United States Merchant Marine Academy, Kings Point, NY, USA\\
$^{2}$National Astronomical Observatory of Japan, Mitaka, Japan\\
$^{3}$Space Science Institute, 4750 Walnut St, Suite 205, Boulder, CO,80301, USA\\
$^{4}$School of Physical Sciences, The Graduate University for Advanced Studies, Shonankokusaimura, Hayama, Miura District, Kanagawa 240-0193, Japan\\
$^{5}$Astronomical Observatory of Jagiellonian University, Krakow, Poland\\
$^{6}$Jagiellonian University, Doctoral School of Exact and Natural Sciences, Krakow, Poland\\
$^{7}$Finnish Centre for Astronomy with ESO (FINCA), University of Turku, Finland\\
$^{8}$Department of Mathematics, University of Wroclaw, Poland\\
$^{9}$Department of Statistics, Lund University, Sweden\\
}

\date{Accepted XXX. Received YYY; in original form ZZZ}

\pubyear{2023}

\begin{document}
\label{firstpage}
\pagerange{\pageref{firstpage}--\pageref{lastpage}}
\maketitle

% Abstract of the paper
\begin{abstract}

Classifying Active Galactic Nuclei (AGN) is a challenge, especially for BL Lac Objects (BLLs), which are identified by their weak emission line spectra. 
To address the problem of classification, we use data from the $4^{th}$ \textit{Fermi} Catalog, Data Release 3. 
Missing data hinders the use of machine learning to classify AGN. 
A previous paper found that Multiple Imputation by Chain Equations (MICE) imputation is useful for estimating missing values.
%Removed a reference to Gibson 2022 above as requested by Editor. 
Since many AGN have missing redshift and the highest energy, we use data imputation with MICE and K-nearest neighbor (kNN) algorithm to fill in these missing variables. 
Then, we classify AGN into the BLLs or the Flat Spectrum Radio Quasars (FSRQs) using the SuperLearner, an ensemble method that includes several classification algorithms like logistic regression, support vector classifiers, Random Forests, Ranger Random Forests, multivariate adaptive regression spline (MARS), Bayesian regression, Extreme Gradient Boosting.
We find that a SuperLearner model using MARS regression and Random Forests algorithms is 91.1\% accurate for kNN imputed data and 91.2\% for MICE imputed data.
Furthermore, the kNN-imputed SuperLearner model predicts that 892 of the 1519 unclassified blazars are BLLs and 627 are Flat Spectrum Radio Quasars (FSRQs), while the MICE-imputed SuperLearner model predicts 890 BLLs and 629 FSRQs in the unclassified set. 
Thus, we can conclude that both imputation methods work efficiently and with high accuracy and that our methodology ushers the way for using SuperLearner as a novel classification method in the AGN community and, in general, in the astrophysics community.

%Measuring redshift for Active Galactic Nuclei (AGN) is a challenge, especially for BL Lac Objects (BLLs), which are identified by their weak emission line spectra. Missing data hinders the use of machine learning to classify AGNs. In a previous paper, \cite{Gibson2022} found that MICE imputation is useful for estimating missing values. We expand upon that research using data from the $4^{th}$ Fermi Catalog, Data Release 3. We build an ensemble classification model using the SuperLearner library in R. We start by evaluating the effect of missing data on Logistic Regression, then we evaluate various classification algorithms, logistic regression, support vector classifiers, random forests, Ranger random forests, MARS regression, Bayesian regression, extreme gradient boosting, and Neural Nets that we will then add to SuperLearner. Finally, we evaluate parallel analysis for kNN and MICE imputation. We find that a SuperLearner model using Logistic Regression, Support Vector Classifiers, and Random Forests algorithms are 92.0\% accurate for kNN imputation and 92.6\% accurate for MICE Imputed data. Furthermore, the KNN-imputed SuperLearner model predicts that 738 of the 1312 unclassified blazars are BLLs and 574 are Flat Spectrum Radio Quasars (FSRQs), while the MICE-imputed SuperLearner model predicts 774 BLLs and 538 FSRQs in the unclassified set. 
\end{abstract}

% Select between one and six entries from the list of approved keywords.
% Don't make up new ones.

%%%%%%%%%%%%%%%%%%%%%%%%%%%%%%%%%%%%%%%%%%%%%%%%%%

%%%%%%%%%%%%%%%%% BODY OF PAPER %%%%%%%%%%%%%%%%%%

\section{Introduction}

 Blazars are the extreme class of Active Galactic Nuclei (AGN) with jets oriented close to the observer's line of sight \citep{Liodakis2016}. They are traditionally composed of two main classes, BL Lacertae Objects (BLL) and Flat Spectrum Radio Quasars (FSRQ). The main distinction between these classes of objects is that BLLs often show no or very weak emission line spectra, whereas FSRQs typically show broad emission lines. The origin of the different blazar classes is still a matter of debate. However, it is often attributed to either a difference in accretion modes \cite[e.g.,][]{ghisellini2011transition} or observational biases \cite[e.g.,][]{Padovani2019}.

The orientation of the jets close to the line of sight causes extreme relativistic and projection effects \citep{Liodakis2017,Liodakis2017-II,Liodakis2018}, challenging the predominant classification scheme \cite[e.g.,][]{kharb2010ApJ...710..764K}. Blazars are the most numerous extragalactic $\gamma$-ray sources detected by the {\it Fermi} $\gamma$-ray space telescope ({\it Fermi}, \citealp{2020ApJS..247...33A}). 
Out of the {\it Fermi} blazars, BLL dominates the population, accounting for more than 40\% of the sources. FSRQs account for only about 20\%, while blazars of uncertain type (BCUs) are about 35\% of all {\it Fermi} blazars \citep{Ajello2022}.
Given this disparity, the $\gamma$-ray characteristics of BLL and FSRQs may be distinct enough to be used as classification diagnostics, as opposed to the relative strength of atomic line spectra, as has been done historically. Human-based classification becomes highly inefficient, especially in multivariate data spaces. 
In this regard, machine learning (ML) is essential to detect patterns that otherwise would be unseen.
We can broadly divide supervised machine learning into regression and classification. In the former, a value is estimated from the data, while in the latter, the class of an object is predicted based on a data set.  
Previous attempts using machine learning methods have focused on predicting blazar properties such as redshift \cite[e.g.,][]{Dainotti2021,Narendra2022,Gibson2022}, synchrotron peak \cite[eg.,][]{Glauch2022} and classification of {\it Fermi} blazars of uncertain type and unidentified sources \cite[e.g.,][]{Chiaro2016,Liodakis2019,Finke2021,Coronado-Blazquez2022}. 
\cite{Liodakis2019}  combined infrared data and optical polarization observations from the RoboPol survey \citep{Ramaprakash2019,Blinov2021} in Random Forest and logistic regression networks to identify blazars in the unidentified {\it Fermi} uncertainty regions. They verified the candidate AGN proposed by \cite{Mandarakas2019} for 3FGL~J0221.2+2518 and paved the way for future optopolarimetric surveys (e.g., PASIPHAE, \citealp{Tassis2018}) to find more $\gamma$-ray emitting AGNs.
\cite{Dainotti2021}, for the first time, used SuperLearner, an ensemble ML method that leverages the advantages of several constituent ML methods, in the AGN community to derive the redshift of AGN with an accuracy that is comparable to other methods shown in the literature \citep{brescia2013photometric,Brescia2019}.
In the following papers \cite[e.g.,][]{Narendra2022,Gibson2022} we show the reliability of the Multivariate Imputation by Chained Equation (MICE) \citep{van2011mice} in solving the problems of AGN variables missing at random and how these confirm the results of the accuracy on the redshift inference. 
Considering that a larger sample with more complete variables is essential for the classification purposes in the current analysis, we adopted the methods of MICE and SuperLearner for the AGN classification.

This study uses regression to estimate the data missing in the 4th {\it Fermi} catalog \cite[4FGL,][]{2020ApJS..247...33A} and classification to distinguish between BLLs and FSRQs among the BCUs. 
This has already been attempted by other groups in different wavelengths \cite[e.g.,][]{kovacevi2020MNRAS.493.1926K,Bhat2022A&A...660A..87B,Butter2022JCAP...04..023B,Chiaro2021JHEAp..29...40C,sahakyan2022} using single models. 
{\cite{kovacevi2020MNRAS.493.1926K} leverage a single artificial neural network for classifying 1329 BCUs into 801 BLLs and 406 FSRQs, with 122 remaining unclassified.}
{\cite{Bhat2022A&A...660A..87B} use Random Forest and neural network models to classify unassociated \textit{Fermi} catalog objects into pulsars and AGNs.}
\cite{Kang2019ApJ...887..134K} use support vector machines, artificial neural networks, and Random Forests, achieving 91.8\% to 92.9\% accuracy and predict that the unclassified blazars consist of 724 BL BLacs and 332 FSRQs with 332 still remaining unclassified.  
\cite{Butter2022JCAP...04..023B} use a Bayesian neural network for classifying \textit{Fermi} Blazars into BLLs and FSRQs and provide uncertainty estimates in their predictions.
\cite{Chiaro2021JHEAp..29...40C} have improved the artificial neural network used in \cite{kovacevi2020MNRAS.493.1926K} and provided a larger classification result for the 4FGL dataset. 
Therefore, in addition to using only data provided in the 4FGL, we attack this problem by building an ensemble of methods to find the optimal classification model. 

The paper is organized as follows. In section \ref{sec:data}, we present the data used in the analysis. In section \ref{sec:methodology}, we lay down the methodology and build our models. In section \ref{sec:results}, we present our results, and in section \ref{sec:conclusions}, we summarize our conclusions. We also included an appendix, see Sec. \ref{sec:appdx}, where we analyze classification for all AGN classes present, not just blazars.

\section{Data sample}\label{sec:data}

As previously stated, our data sample is taken from 4FGL \citep{2022ApJS..260...53A}. %the 4th Fermi Catalog, Data Release 3 (4FGL) \citep{2020ApJS..247...33A,2022ApJS..260...53A}.
Data consists of 3770 observations of 22 variables. Table \ref{tab:ferm_var} contains a short description of the variables included in this study; a more extensive description of each variable is obtained from \cite{2020ApJ...892..105A}. 

\begin{table*}
    \centering
    \caption{4FGL, see also Table 6 in \protect\cite{2022ApJS..260...53A}}
    \label{tab:ferm_var}
    \begin{tabular}{lll}%{c|c|c}
    \hline
    \hline
    Variable & Units & Description \\
    \hline
     Source Name & - & Based on RA and DEC: 4FGL  JHHMM.m+DDMM. \\
     Class & - & Class designation of likely associated  source. \\
     Association & - &Name of identified or likely associated  source. \\
     Energy Flux 100 & $erg*cm^{-2}*s^{-1}$ & Energy Flux in  from 100 MeV  to 100 GeV.   \\
     Highest Energy & MeV & Energy of the highest-energy photon,  association probability $P > 0.95$\\
     Variability Index & - &Difference between flux per time  interval and average flux.\\
     Fractional Variability & - & Fractional variability computed from  fluxes each year.\\
     Gaia G Magnitude & - & G-band magnitudes from the Gaia  survey.\\
     $\nu_{syn}$ & $s^{-1}$ & Synchrotron-peak frequency in  observer frame. \\
     $\nu F \nu_{syn}$ & $erg*cm^{-2}*s^{-1}$ & Spectral energy distribution at  synchrotron-peak frequency\\
     Spectrum Type & - & The spectral type (PowerLaw,  LogParabola, PLSuperExpCutoff). \\
     PL Index & - & Photon index for the PowerLaw fit. \\
     LP Index &- & Photon index at pivot energy for  LogParabola fit.\\
     LP Beta &- & Curvature parameter for LogParabola  spectrum type.\\
     LP SigCurv & $\sigma$ units & Significance of the improvement  between Power Law and  Log Parabola fits. \\
     LP Epeak & MeV & Peak energy in $\nu F_{\nu}$ estimated from  the Log Parabola model. \\
     PLEC IndexS & - & Photon Index at Pivot Energy when  fitting with PLSuperExpCutoff. \\
     PLEC ExpfactorS & - & Spectral Curve at Pivot Energy when  fitting with PLSuperExpCutoff. \\
     PLEC Exp Index & - & Exponential index when fitting  with LSuperExpCutoff. \\
     PLEC SigCurv & $\sigma$ units &   Significance of the improvement  for the PLSuperExpCutoff   model to PL. \\
     PLEC EPeak & MeV & Peak energy in $\nu F_{\nu}$ estimated  for the PLSuperExpCutoff model. \\
     Redshift & - & Redshift of identified or likely  associated source.\\
    \end{tabular}
    
\end{table*}

Data includes 18 numerical variables related to  $\gamma$-ray emission, variability, optical magnitude, and spectral characteristics. Table \ref{tab:nds} contains the descriptive statistics for these measurements for the blazar classes, BLL, FSRQ, and BCU. 
The data also includes two categorical variables, classification, and spectrum type. The data contains 2192 Power Law (PL) spectrum sources, 1572 Log Parabola spectrum sources, and 6 PL SuperExponential Cutoff sources (see \cite{Ajello2022}). We created "is{\_}PL" as a categorical variable where blazars with a Power Law spectrum have is{\_}PL of one, and those with other spectral distributions have is{\_}PL of zero.
Note that in this sample, we have missing data in twelve features: highest energy,  $\nu_{syn}$, $\nu F_{\nu}$, Gaia G magnitude, the significance between the PL and the Log parabola spectrum, (LP SigCurv), the Peak energy in the $\nu_{syn}$, $\nu F_{\nu}$ spectrum (LP EPeak), the photon index at the pivot energy when fitting with the super-exponential Cutoff (PLEC IndexS), the spectral Curve at pivot energy when fitting with the super-exponential cutoff (PLEC ExpfactorS), the exponential index when fitting with the exponential cutoff (PLEC Exp Index), the significance of the improvement of the superexponential fit to the power-law fit (PLEC SigCurv), the peak energy in the $\nu_{syn}$, $\nu F_{\nu}$ spectrum estimated for the PL SuperExponential Cutoff model (PLEC EPeak), and redshift. Since  Peak Flux contained 2210 missing entries, we dropped this feature from the analysis. 

\begin{table*}
    \centering
    \caption{Numerical Data Summary}
    \label{tab:nds}
    \begin{tabular}{llllllll}
  \hline
  \hline
 & Min. & 1st Qu. & Median & Mean & 3rd Qu. & Max & NA's \\ 
  \hline
    Energy Flux100 & 3.708e-13 & 1.804e-12 & 3.128e-12 & 8.045e-12 & 6.314e-12 & 8.440e-10 &  \\ 
    Highest energy &  4.153    &  10.000 & 21.344 &  47.728   &  53.544   & 912.930   & 657   \\ 
    Variability Index & 1.51   &  12.38  & 19.96 &   237.52   &    49.52   & 84340.76   &  \\ 
    Frac Variability & 0.0000 & 0.0000 & 0.3514   & 0.4076   & 0.6122   & 2.9100   &  \\ 
    Gaia G Mag. & 12.81   & 17.79   & 18.58   & 18.56   & 19.46   & 21.45   & 1691   \\ 
    $\nu_{syn}$ & 0.000e+00   & 0.000e+00   & 1.000e+13   & 1.549e+16   & 1.995e+14   & 1.120e+19   &  633 \\ 
    $\nu F_{\nu}$ syn & 0.000e+00   & 0.000e+00   & 7.600e-13   & 2.203e-12   & 1.990e-12   & 2.200e-10   & 633 \\ 
     PL Index & 1.419   & 1.992   & 2.213   & 2.219   & 2.444   & 3.153   &  \\ 
     LP Index & -0.04898   &  1.89016   &  2.13492   &  2.13409   &  2.39093   &  3.20418   &  \\ 
     LP beta & -0.11616   &  0.04101   &  0.09494   &  0.14702   &  0.17660   &  1.00000   &  \\
     LP SigCurv & 0.0000 & 0.8402 & 1.6431 & 2.2372 & 2.8842 & 42.6642 & 27 \\
     LP Epeak & 0 & 187 & 871 & 160352 & 5904 & 111488184 & 567\\
     PLEC IndexS & -0.1597 & 1.8404 & 2.1006 & 2.0863 & 2.3454 & 3.2123 &  27 \\
     PLEC ExpfactorS & -0.10175 & 0.04355 & 0.10634 & 0.22794 & 0.22249 & 5.15443 & 27  \\
     PLEC Exp Index & 0.2419 & 0.6667 & 0.6667 & 0.6662 & 0.6667 & 0.6667 & 27 \\
     PLEC SigCurv & 0.0000 & 0.8771 & 1.7415 & 2.2767 & 2.8841 & 43.4651  &  27 \\
     PLEC EPeak & 0 & 802 & 4691 & 26337 & 16196 & 4593440 & 1526 \\
     Redshift & 0.0000   & 0.2548   & 0.6330   & 0.8577   & 1.2740   & 6.4430   & 1901   \\ 
    \hline
    \end{tabular}
\end{table*}

%Blazars are denoted by the equivalent width of resonant emission lines in their optical spectra. 
%Sources with broad emission lines are classified as Flat Spectrum Radio Quasars, whereas sources with weak or no emission lines are classified as BLL (BLLs). The data also contained blazars of unclassified type called BCUs. 
%Measuring the redshift (z) of blazars has been a cumbersome and observationally expensive endeavor. 
%The situation is further complicated by the absence of emission lines in the most numerous class of $\gamma$-ray loud blazars, i.e., BLL. 

\begin{figure*} 
\centering
% \subfloat[Injection Wind Slow Cooling $\nu_m < \nu < \nu_c$]{\label{fig1_a}

    \includegraphics[width=0.68\columnwidth]{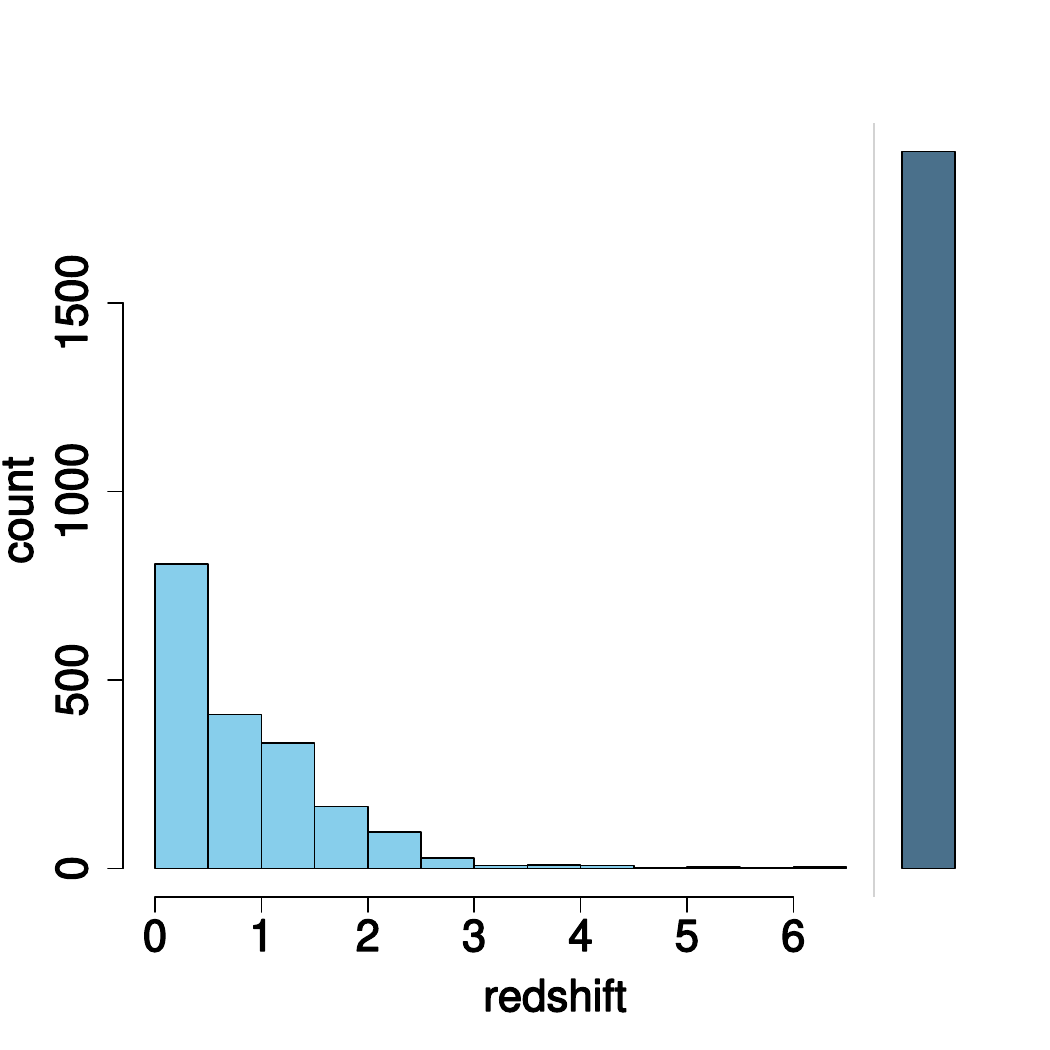} 
    %\caption{Table \ref{tab:4} Injection Wind Slow Cooling $\nu_m < \nu < \nu_c$, Table \ref{tab:5} Injection Wind Slow Cooling $\nu_a < \nu < \nu_c$} &
    %\includegraphics[width=0.45\textwidth]{figures/T2.pdf} \\
    % \caption{Table \ref{tab:4} No Injection ISM Slow Cooling $\nu_m < \nu < \nu_c$, Table \ref{tab:5} No Injection ISM Slow Cooling $\nu_a < \nu < \nu_c$}\
    \includegraphics[width=0.68\columnwidth]{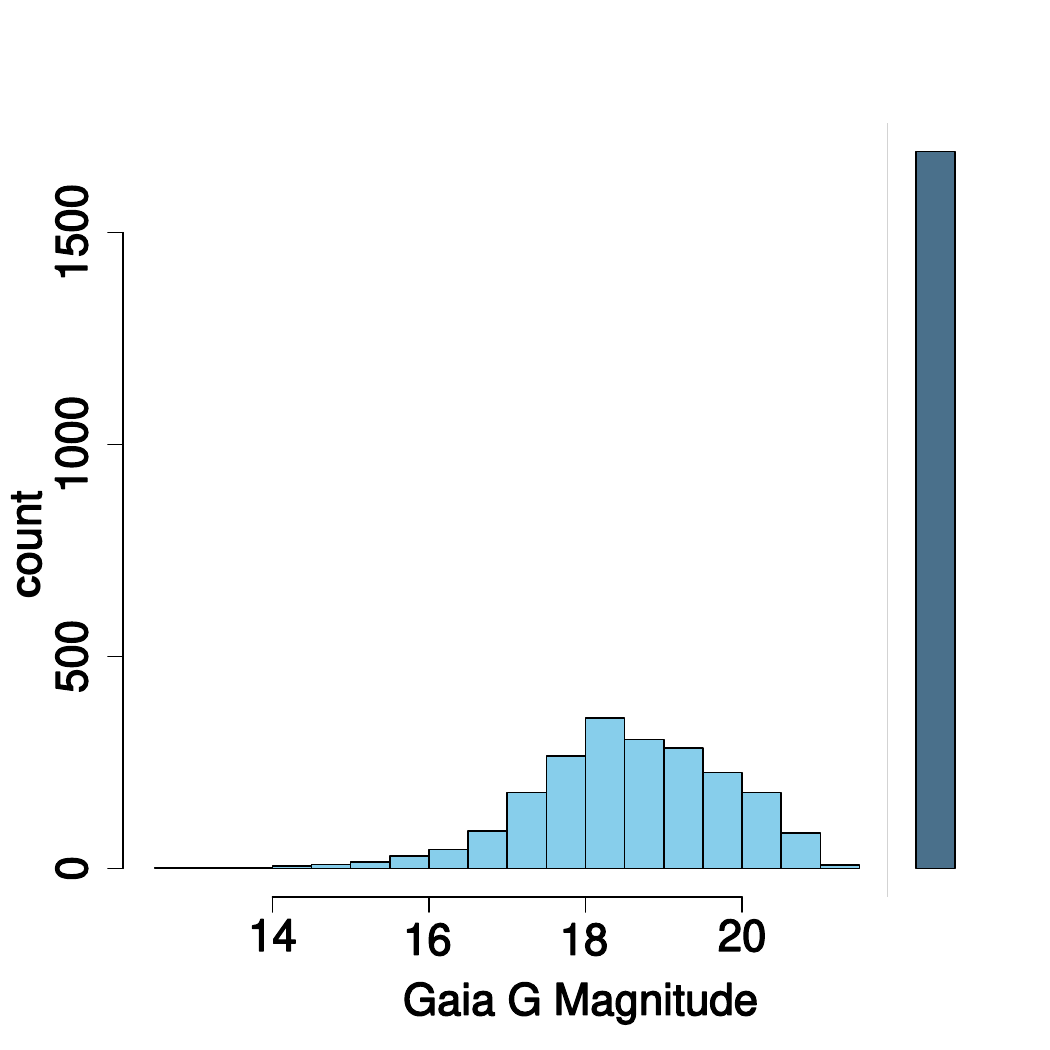}
    % \caption{Table \ref{tab:4} No Injection Wind Slow Cooling $\nu_m < \nu < \nu_c$\\
    % Table \ref{tab:5} No Injection Wind Slow Cooling $\nu_a < \nu < \nu_c$}
    \includegraphics[width=0.68\columnwidth]{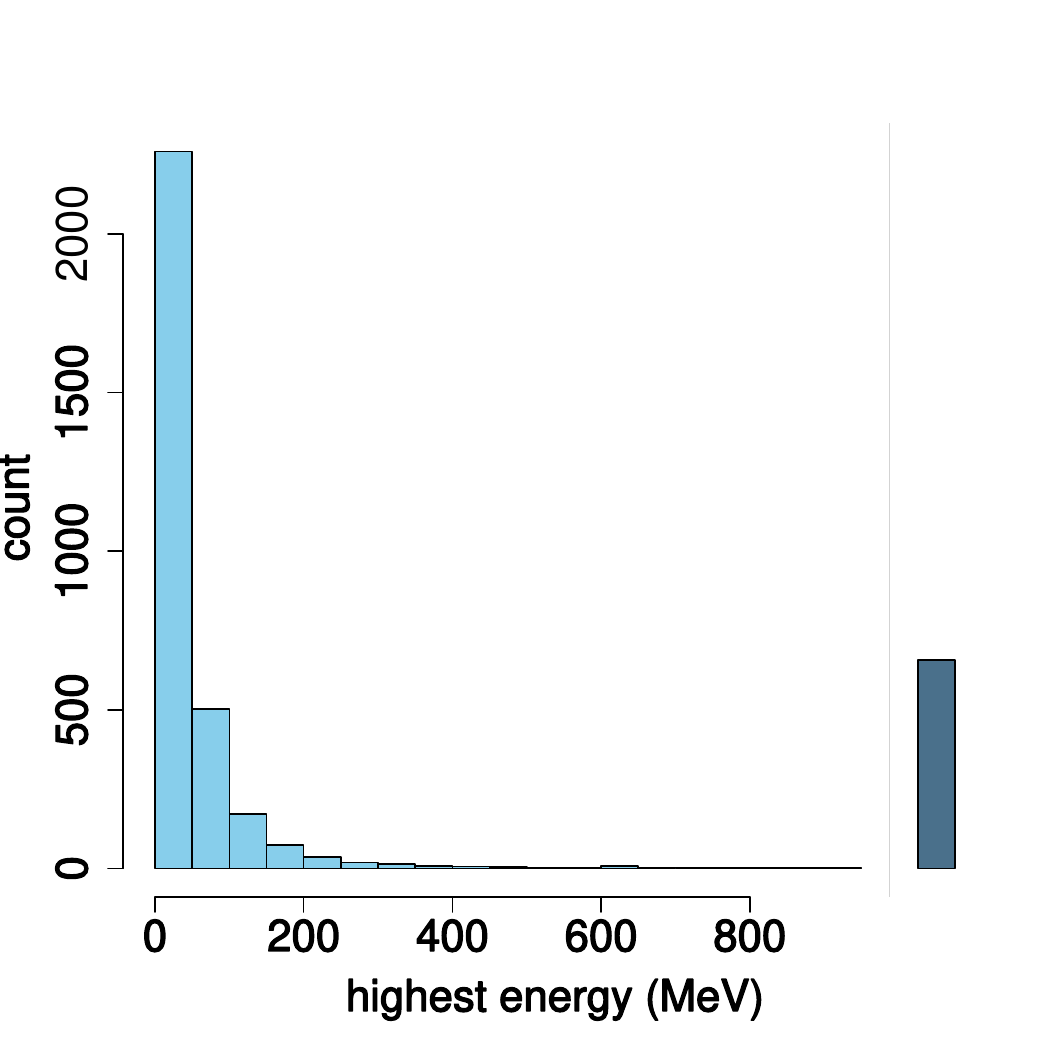}
    % \caption{Table \ref{tab:6} Injection Wind $\nu_m < \nu < \nu_c$\\
    % Table \ref{tab:7} Injection Wind $\nu_a < \nu < \nu_c$}
    \includegraphics[width=0.68\columnwidth]{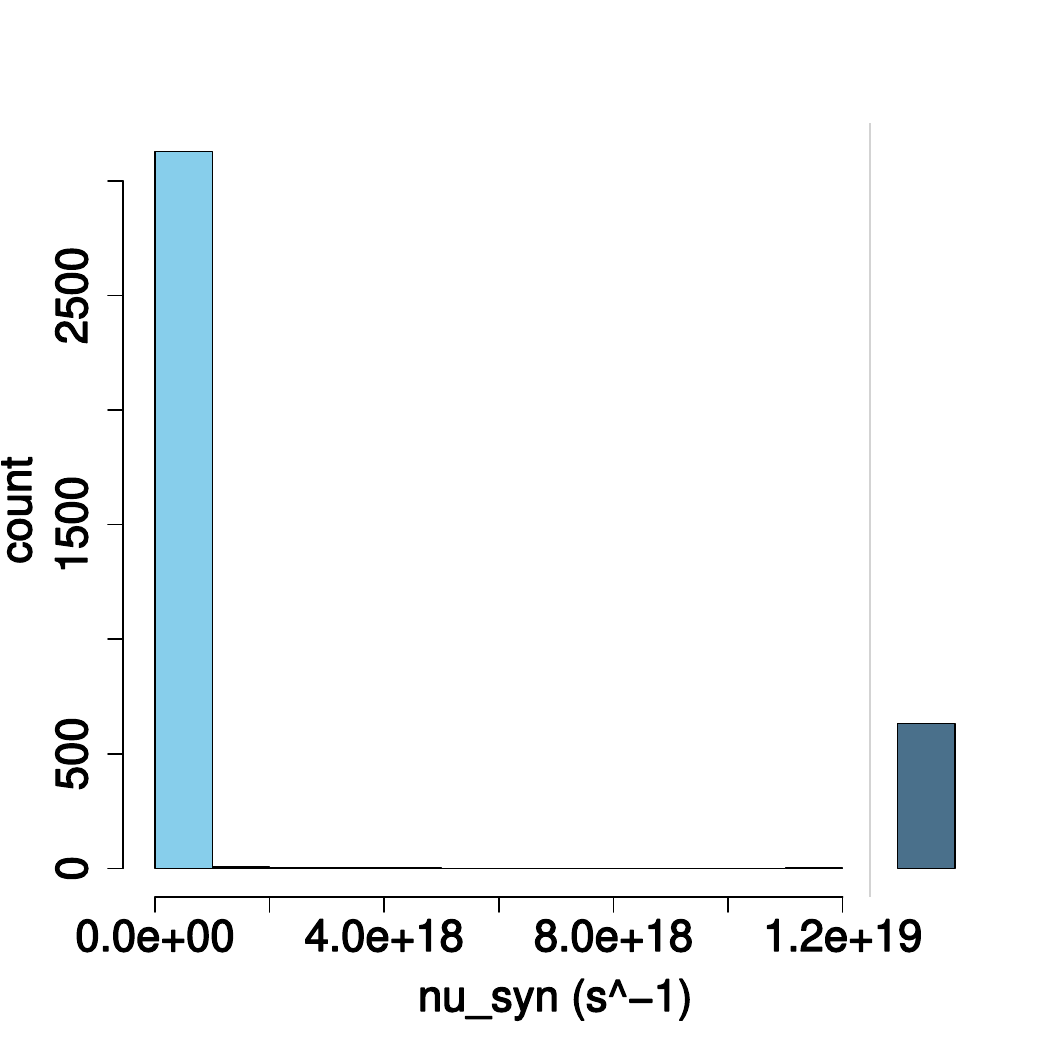}
    \includegraphics[width=0.68\columnwidth]{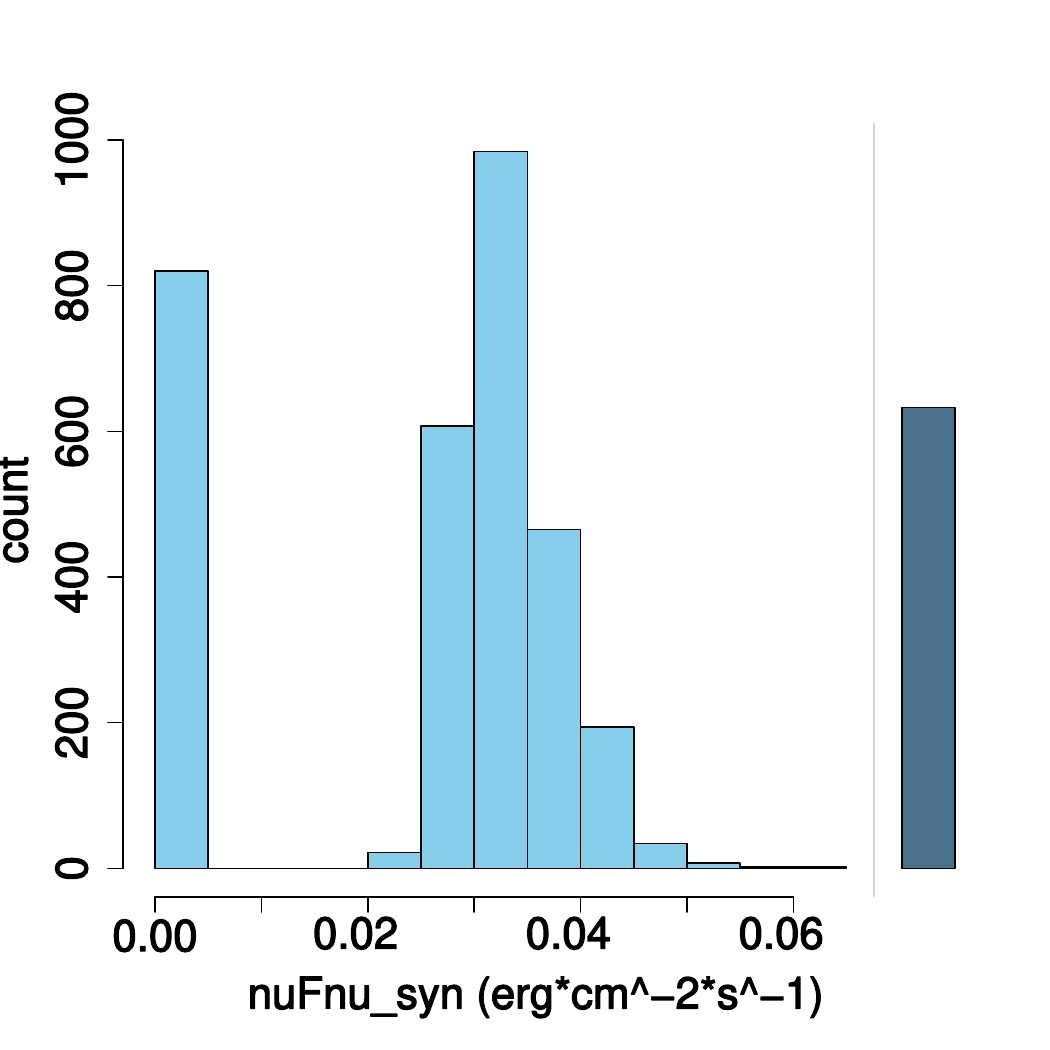}
    % \caption{Table \ref{tab:6} No Injection ISM $\nu > \nu_c$\\
    % Table \ref{tab:7} No Injection ISM $\nu > \nu_c$}
    \includegraphics[width=0.68\columnwidth]{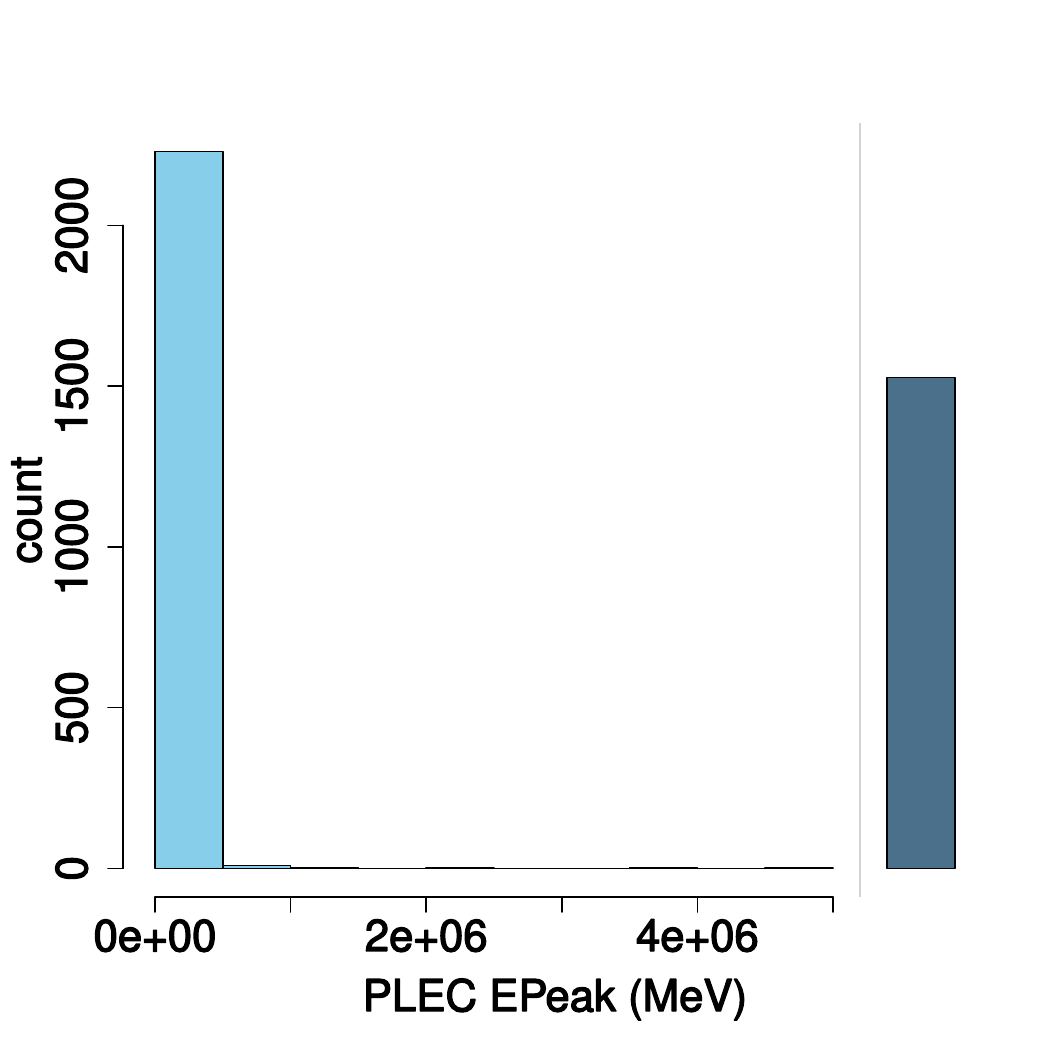}
    % \caption{Table \ref{tab:6} No Injection ISM $\nu_m < \nu < \nu_c$\\
    % Table \ref{tab:7} No Injection ISM $\nu_a < \nu < \nu_c$}
    \includegraphics[width=0.68\columnwidth]{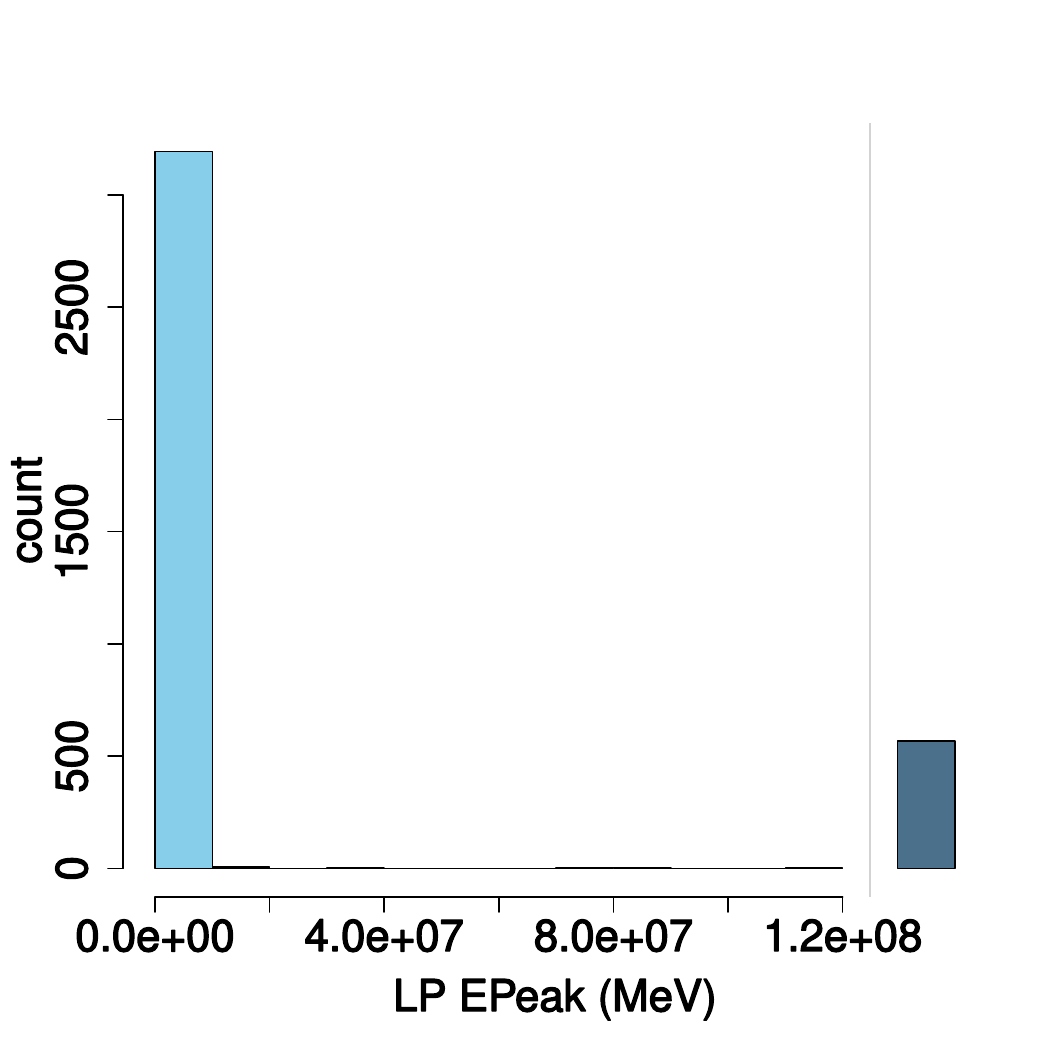}
% \caption{Table \ref{tab:6} No Injection Wind $\nu > \nu_c$\\
% Table \ref{tab:7} No Injection Wind $\nu > \nu_c$}
\caption{ 
Top left: Redshift distribution of the blazar set, including the scale of the missing data.
Top middle: Gaia G Magnitude distribution of the blazar set, including the scale of the missing data.
Top right: Highest Energy (MeV) distribution of the blazar set, including the scale of the missing data.
Middle left: $\nu_{syn}$ ($s^{-1}$) distribution of the blazar set, including the scale of the missing data.
Middle: $\nu F \nu_{syn}$ ($erg*cm^{-2}*s^{-1}$) distribution of the blazar set, under $8^{th}$ root transformation, including the scale of the missing data.
Middle right: PLEC EPeak (MeV) distribution of the blazar set, including the scale of the missing data.
Bottom: Log Parabola EPeak (MeV) distribution of the blazar set, including the scale of the missing data.
}
\label{fig:missing}
\end{figure*}

%We have calculated the descriptive statistics of the data using the summary function in R. These statistics are shown in Table \ref{tab:nds}. %For numerical data, summary calculates the minimum, 1st quartile, median, mean, 3rd quartile, and maximum, it also gives the number of missing observations. For categorical data, which are called factors in R, the summary function gives a count of each category. %Histograms and box-and-whisker plots were created using the ggplot2 library, 
We show histograms to compare distributions of missing data for features with more than 27 missing entries,
see Fig. \ref{fig:missing}.
%see Fig. \ref{fig:red1} through \ref{fig:LPE1}. 

\section{Methodology}\label{sec:methodology}
The methodology is mainly divided into two steps: 
first, we treat the missing data and secondly, we evaluate the classification through the ensemble method of SuperLearner \citep{van2007super,polley2010super}; including the accuracy of the individual machine learning models outside of the SuperLearner ensemble. 
The methodology is described below in the following subsections.

\subsection{Data Handling}\label{sec:handling}

%We use the $3^{rd}$ data release of the $4^{th}$ Fermi Catalog. 
As this study focuses on the classification of blazars, we did not include non$-$blazars from the data set. The total blazars are 1457 BLLs, 794 FSRQs, and 1519 BCUs.
We divided these data into two sets, 2251 classified blazars for the training and testing of the various models we use, and 1519 BCUs for evaluation. 
However, each set contains missing data. 

The set of classified blazars (BLLs and FSRQs) contains 927 missing PLEC EPeaks, 638 missing Gaia G Magnitudes, 607 missing redshifts, 301 missing LP EPeaks, 215 missing Highest Energies, 201 missing for both $\nu_{syn}$ and  $\nu F \nu_{syn}$, and one missing each for LP SigCurv, PLEC IndexS, PLEC ExpfactorS,  PLEC Exp Index, and PLEC SigCurv. 

\begin{figure*}
\centering
\includegraphics[width=0.9\textwidth]{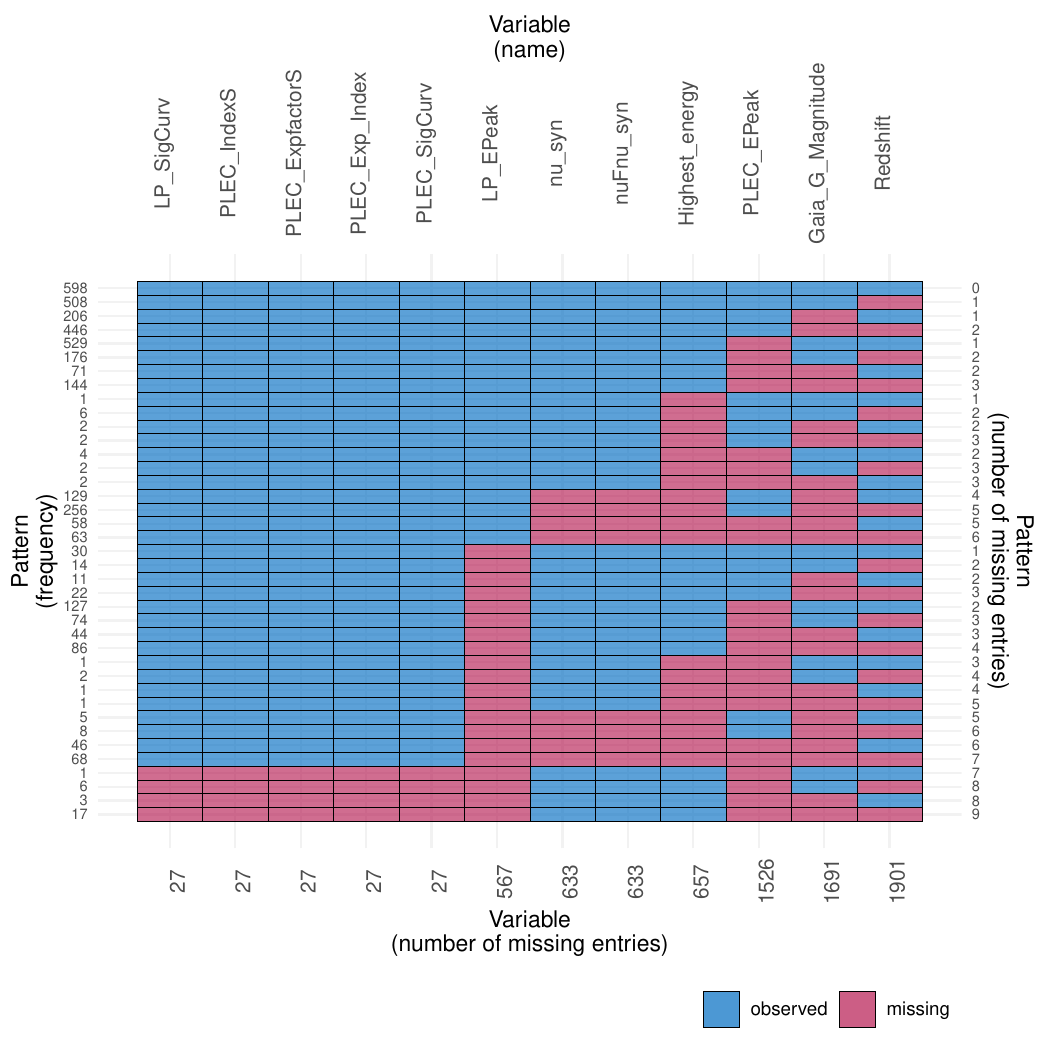}
\caption{
This figure shows all the missing data. 
The blue boxes show the complete data, and the red boxes show the missing data. 
The variables' names are presented at the top of the figure, while the number at the bottom of the figure represents the number of missing entries for each variable.
The number on the right shows the number of variables with missing entries, and the corresponding number on the left shows the number of AGNs that have missing entries in those variables.
}
\label{fig:MISS}
\end{figure*}

The set of unclassified blazars (BCUs) contains 1294 missing redshifts, 1053 missing Gaia G Magnitudes, 599 missing PLEC EPeaks, 442 missing Highest Energies, 432 missing for both $\nu_{syn}$ and  $\nu F \nu_{syn}$, 266 missing LP EPeaks, and 26 each for LP SigCurv, PLEC IndexS, PLEC ExpfactorS, PLEC Exp Index, and PLEC SigCurv.
We show the missing entry compared to the full set of variables in Fig. \ref{fig:MISS}, which can be read as follows:
for example, for the variable LP\_EPeak, 567 AGNs have missing entries. 
The number 567 is mentioned at the bottom of the figure corresponding to LP\_EPeak.
Along the rows of the figure, for example, row 3, it reads as 206 AGNs having missing data in two variables. 
These two variables are Gaia G Magnitude and Redshift.
We discuss our methods for handling these missing data below.

\subsection{Missing Data}

{Here we have applied two methods for the imputation of the missing data. These are: }
%We used three methods.
\begin{itemize}
\item
Logistic regression: to form a basis of comparison for the effect of the missing data, we have trained a logistic regression (logit) model by not using the measurements with missing data. Logistic regression calculates the log$-$odds of a binary outcome, the response variable, by one or more predictor variables.
Log$-$odds are defined by:

\begin{equation}
\ln(\frac{p}{1-p})
\end{equation}

where $p$ is the probability (0 to 1) of the binary outcome. 
Here, the binary outcome is a BLL or an FSRQ, modeled by a variable that is 1 for BLLs and 0 for FSRQs. 
In the case of multiple predictor variables, not all will statistically impact the outcome, and thus we perform a back-elimination to remove such variables. 

Back-elimination begins with all variables included in the calculation, and the model contains a significance coefficient, p-value (P), for each variable. 
In statistics, a P=0.05 is the conventional threshold for statistical significance. 
We used the same convention here.
First, we identify the variable with the largest p-value (weakest predictor) and remove it if P $>$ 0.05. 
Then, the model is retrained, and the process is repeated, removing the consecutive weakest predictor with the P $>$ 0.05.
This process is repeated until only variables with p-values smaller than 0.05 (95\% certainty that the coefficient is different from zero) remain.

Regarding the logistic regression models, we have treated the missing data in two ways, by dropping sources (rows) that contain missing data and by dropping the twelve features (columns) with missing data, see Figure \ref{tab:nds}. There are 288 blazars out of 2251 that have complete data. In the sample where rows with missing data are dropped, there are 160 BLLs and 116 FSRQs, for an imbalance ratio (IR) of 1.38. The complete set of identified blazars contains 1457 BLLs and 794 FSRQs for an IR of 1.84. Dropping the columns does not affect the IR. 

In our logistic regression analysis, all features remain significant for the set of 288 classified blazars with complete data. 
Only 12 BCUs have complete data. %the logistic regression model predicts 10 are FSRQs and 2 are BLLs. 
%The logistic regression model trained on only features with complete data predicts that out of the 1519 BCUs, 896 are BLLs, and 623 are FSRQs for an IR of 1.44. 
The complete features that remained significant after back$-$elimination are the Variability Index, Fractional Variability, Spectrum type quantified as 1 for Power Law and 0 for not Power Law, and the Power Law index.

\item
%The method we used to mitigate the effect of the missing data was imputation; 
MICE and kNN: We estimated the missing data using imputation. 
We used two imputation methods, k-nearest neighbors (kNN) \citep{fix1951discriminatory} and Multivariate Imputation by Chained Equation (MICE) \citep{van2011mice,luken2021missing,Gibson2022}. 
%to mitigate the effects of missing data. 
We use MICE since we have already successfully used it in previous analysis \citep{Gibson2022}, and then for comparison, we used the Logistic regression, but we could have also used other models. Although we could try different approaches, this is beyond the scope of the current analysis. 
MICE uses a regression method on the present data to estimate a value for the missing data. 
Both are regression methods that calculate a numerical value based on the data provided. 
kNN estimates a value by finding the k number of closest observations with complete data to the observations with missing data.  
The VIM (Visualization and Imputation of Missing Values) library provided the kNN imputation function for this study. 
The VIM kNN algorithm uses the \cite{Gower_71} coefficient of similarity, which allows the measurement of the similarity of items with mixed numeric and non-numeric data. The default k-value for kNN is equal to 5. 
%is the inner product of the observation of interest the, $k^{th}$ $d(i,j)$, with the other observations of the set $\delta_{i,j,k}$ normalized by range:    
%$$d(i,j) = \frac{\sum_k{\delta_{i,j,k}*d_{i,j,k}}}{\sum_k{\delta_{i,j,k}}}$$ 
To replace a missing variable for a given object, VIM kNN identifies its 5 nearest neighbors according to the Gower similarity coefficient. It then averages the measurement of interest in the 5 neighbors and assigns the average to the missing value in the incomplete observation.

%MICE is also a missing data imputation library for R. 

 The MICE algorithm uses a range of statistical techniques to impute an incomplete column (the target column) by generating 'plausible' synthetic values given to other columns in the data.  
 In this work, we selected the Classification and Regression Tree (“CART”) option to predict unknown values. 
 The MICE function in R allows the generation of multiple imputations of a given value, which reflects uncertainty in the data (\cite{van2011mice}); see Fig. \ref{fig:MICE}. 
 The $\nu F \nu_{syn}$ feature required transformation to converge on a solution in MICE. We discuss this below in section \ref{sec:trns}.
 %For further analysis, we used the 3rd out of 5 imputed values because the different MICE imputation streams give comparable results.

%The MICE algorithm creates a posterior distribution of incomplete data by progressively creating iterative random samples of the data. We used the Classification and Regression Tree (“CART”) method to estimate unknown values. The MICE function in R runs multiple estimations simultaneously (\cite{van2011mice}); see Fig. \ref{fig:MICE}. 

\begin{figure*}
\centering
        \includegraphics[width=0.68\columnwidth]{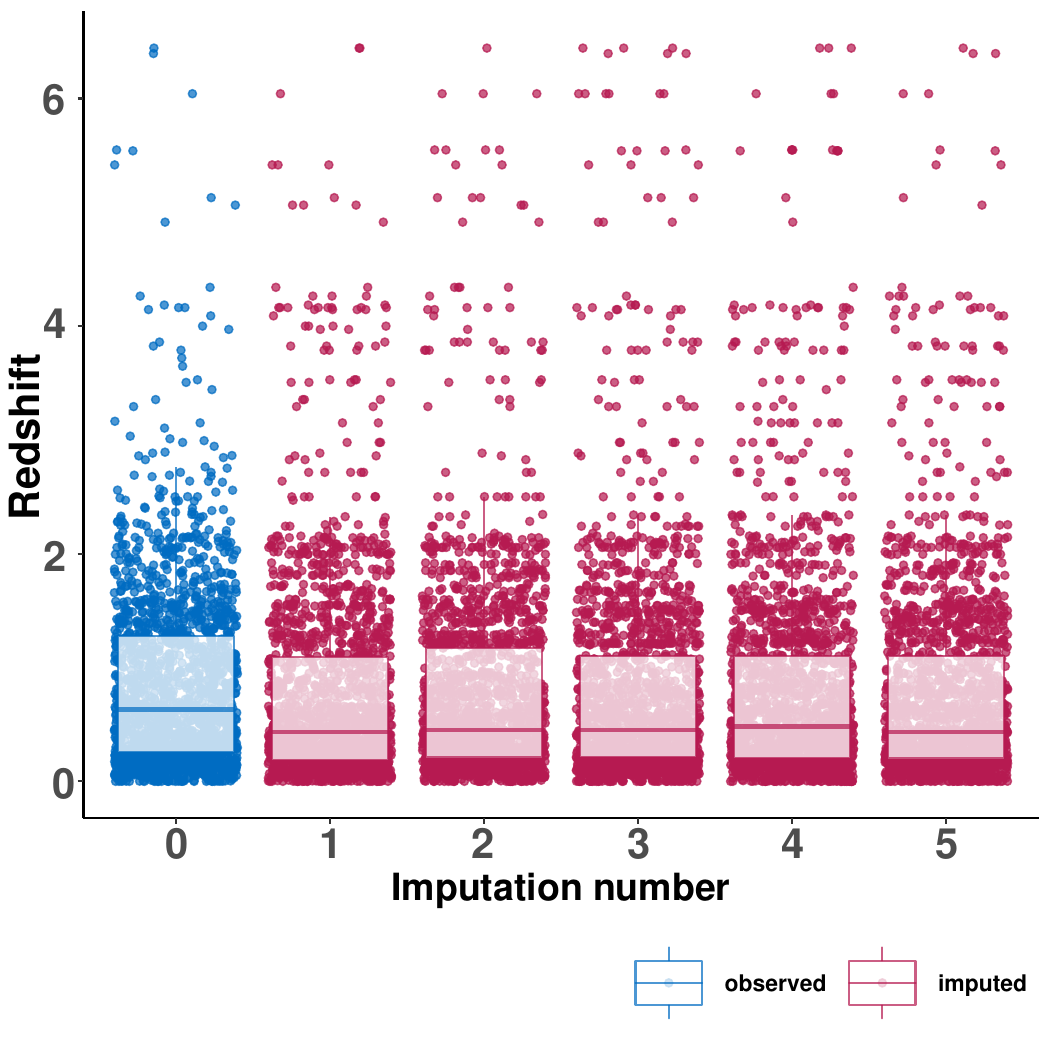}
        \includegraphics[width=0.68\columnwidth]{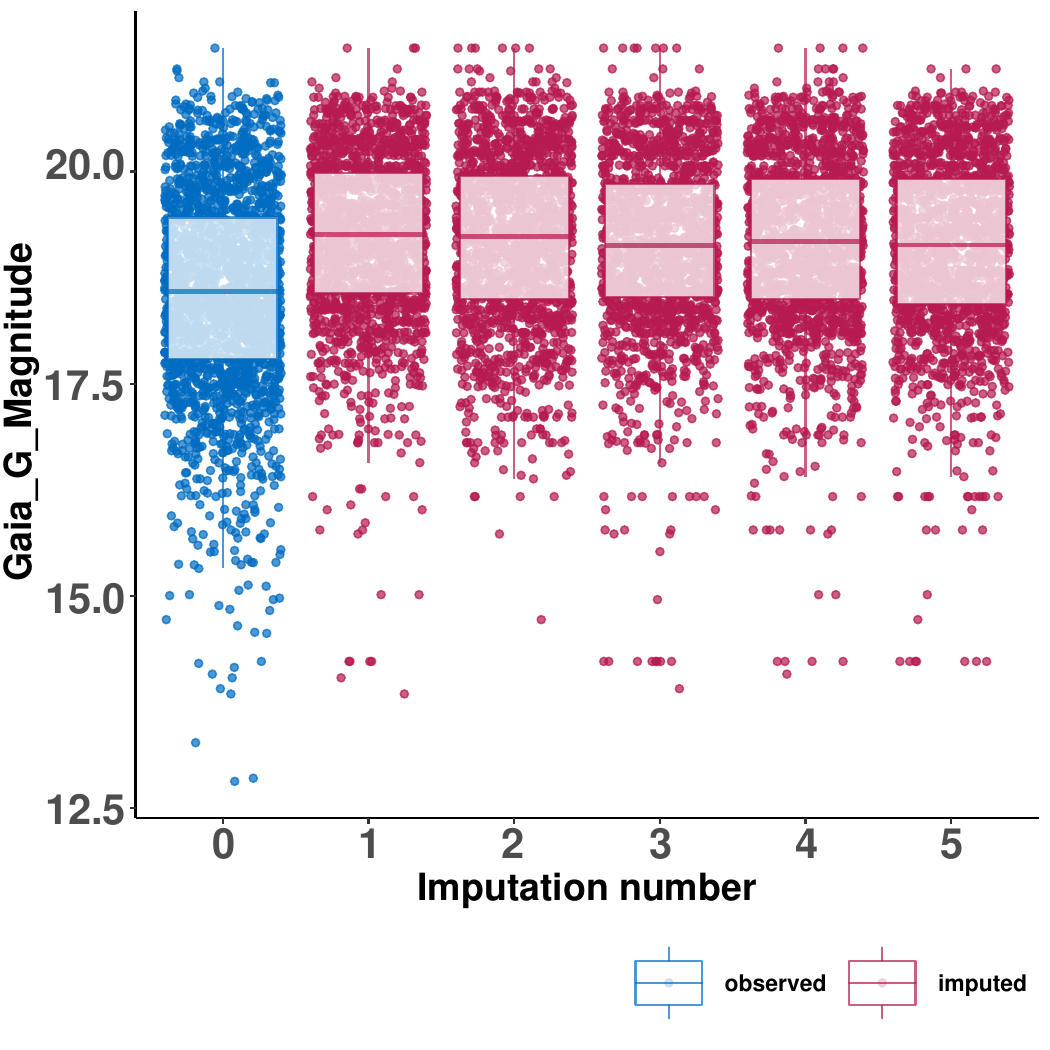}
        \includegraphics[width=0.68\columnwidth]{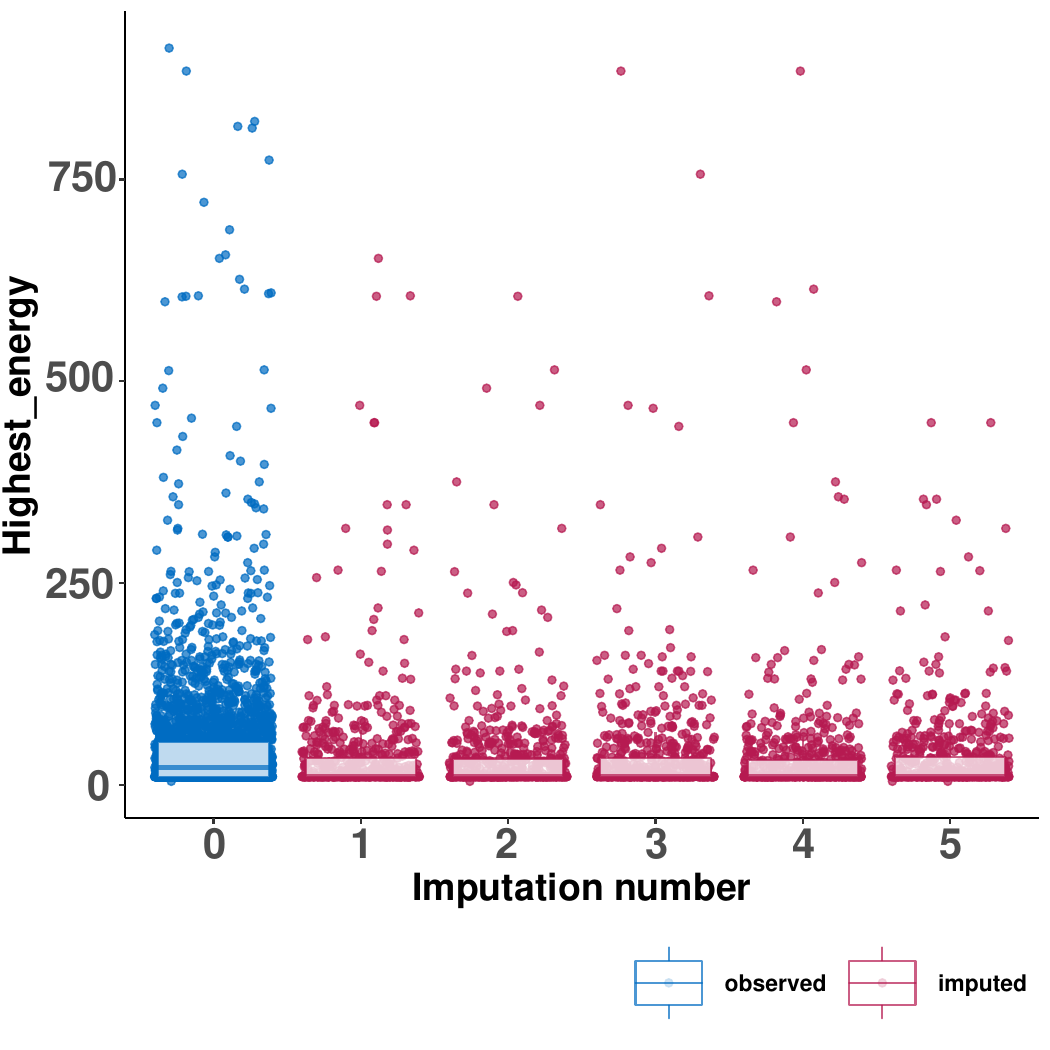}
        \includegraphics[width=0.68\columnwidth]{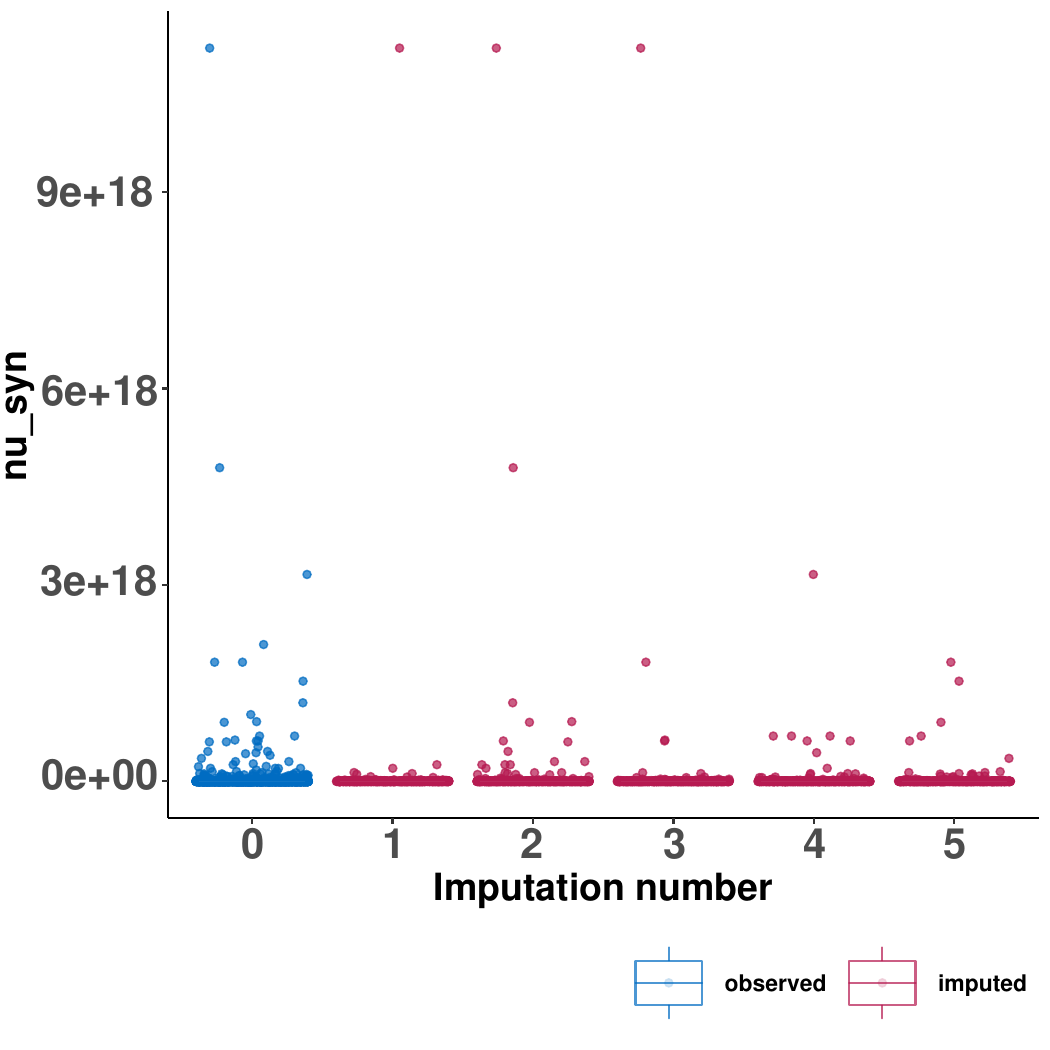}
        \includegraphics[width=0.68\columnwidth]{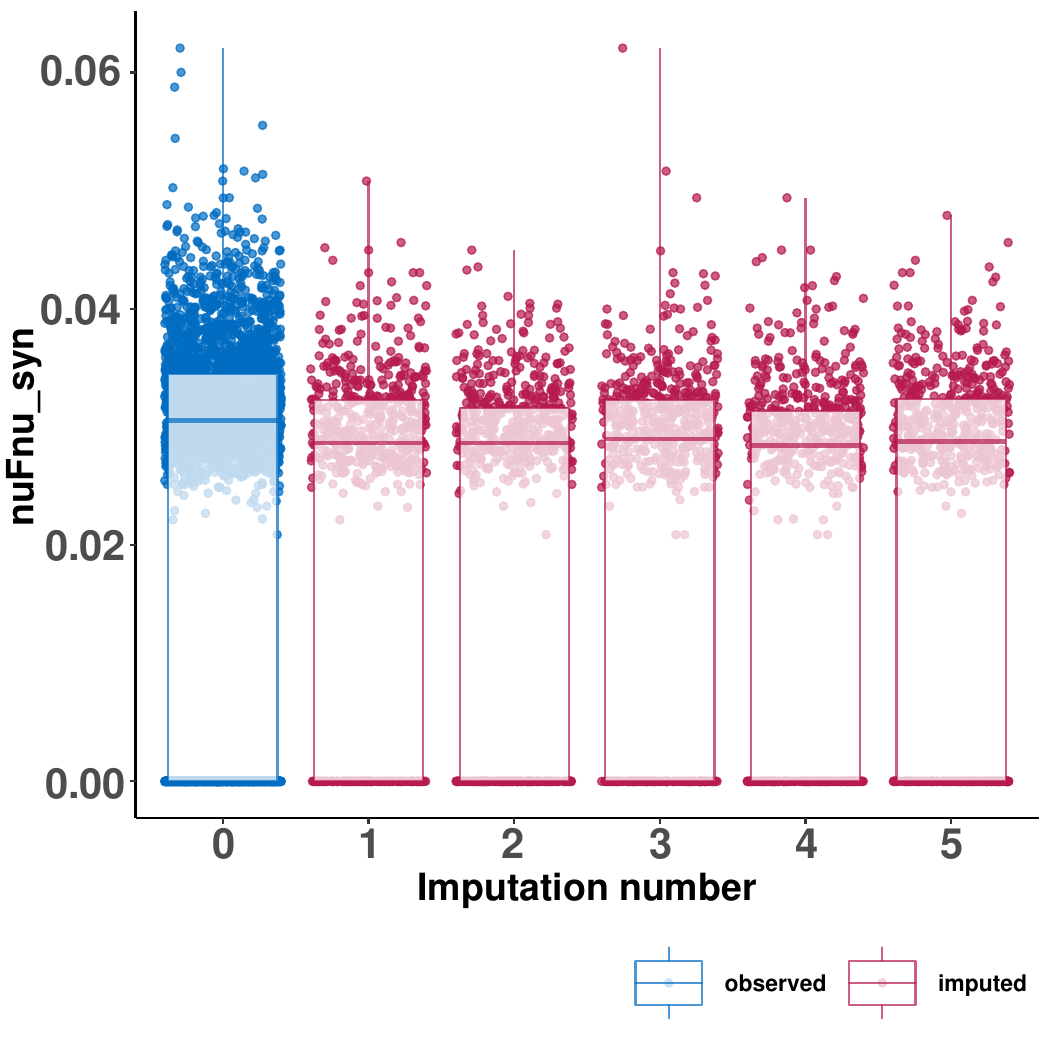}
        \includegraphics[width=0.68\columnwidth]{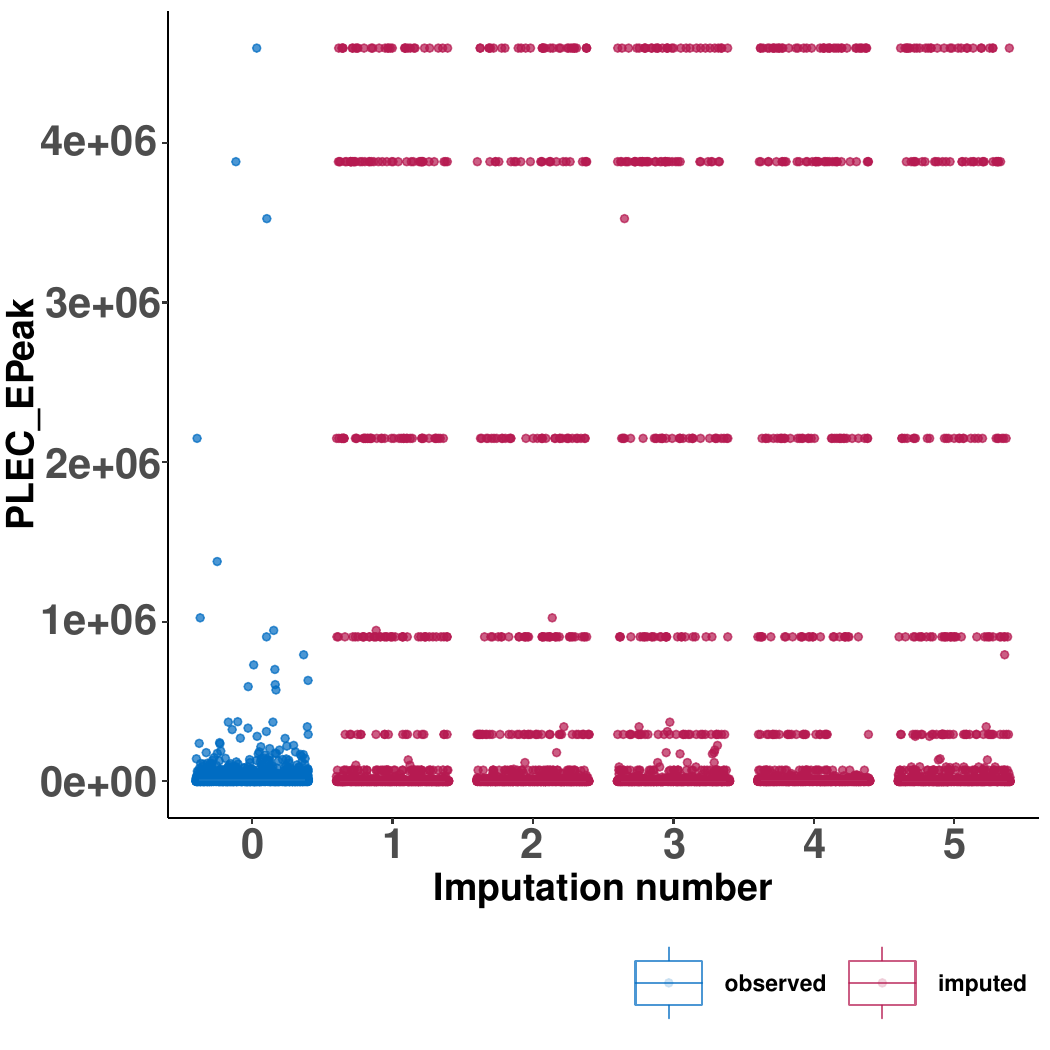}
        \includegraphics[width=0.68\columnwidth]{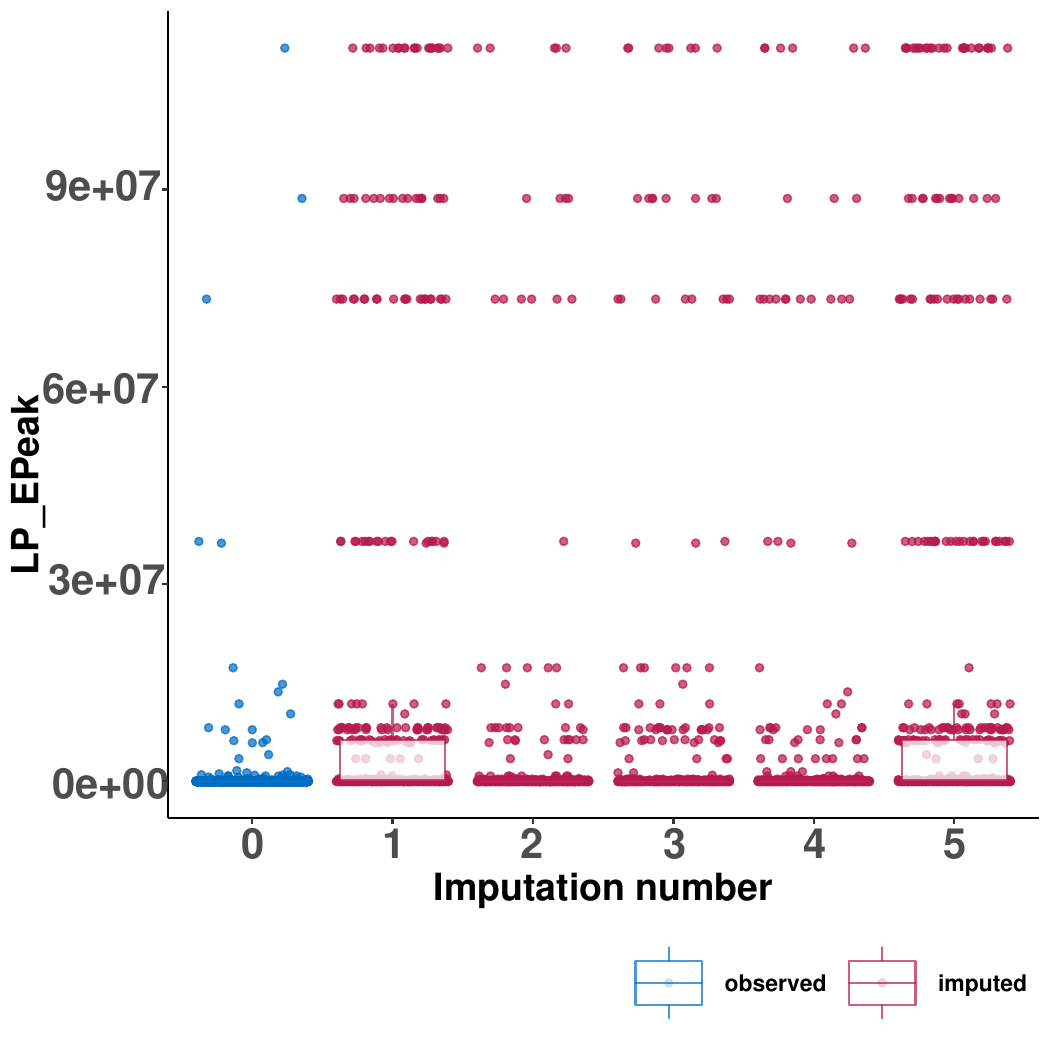}
\caption{
Distribution plots of observed data and the five MICE imputation streams. The Box Plot shows the center 50\% of the distribution with a 75-percentile upper boundary, the middle bar is the median, and the low boundary is the 25-percentile. 
%For subplots \ref{fig:nu2}, \ref{fig:PLEC2}, and \ref{fig:LPE2}, the distribution is such that the box is too small to render
}
\label{fig:MICE}
\end{figure*}

\end{itemize}

\subsection{Data Transformations}\label{sec:trns}
It is always advisable to proceed with the data transformation to improve the predictive power of machine learning. Indeed, in our case,
when performing the MICE imputation, it could not converge on a solution for the $\nu F \nu_{syn}$ feature, which has a bimodal distribution with one mode at zero and one mode distributed at around an order of magnitude of $~ 10^{-12}$. 
A logarithmic transformation is difficult to perform due to the mode at zero. Instead, we performed an 8$^{th}$ root transformation. 
This transformation shifted the second mode to the order of $10^{-2}$, and separated the two modes enough to allow MICE to converge on a solution. 

 Furthermore, we performed a z-score normalization. Certain machine learning algorithms, such as logistic regression, are sensitive to the order of magnitude and may misidentify variables as being non-statistically significant if the units' order of magnitude is much less than other variables in the set. 
 For example, Energy{\_}Flux100 feature has an order of magnitude of $~10^{-12}$.
 Whereas $\nu_{syn}$ has an order of magnitude of $~10^{13}$, as seen in Table \ref{tab:nds}. Machine learning algorithms may ignore Energy{\_}Flux100 based on its low order of magnitude due to the units used and not the underlying physics. The equation for z-score normalization, which mitigates the effect of the order of magnitude differences, is:

\begin{equation}
m_z = \frac{m - \bar{m}}{\sigma_{m}}
\end{equation}

where $m$ is the measurement, $\bar{m}$ is the mean of the measurements, and $\sigma_m$ is the standard deviation. In effect, the z-score is a fractional measure of how many standard deviations an individual measurement varies from the mean. The z-scoring was performed for all numerical data after imputation occurred. 

\begin{figure*}
\centering
\includegraphics[width=2.1\columnwidth]{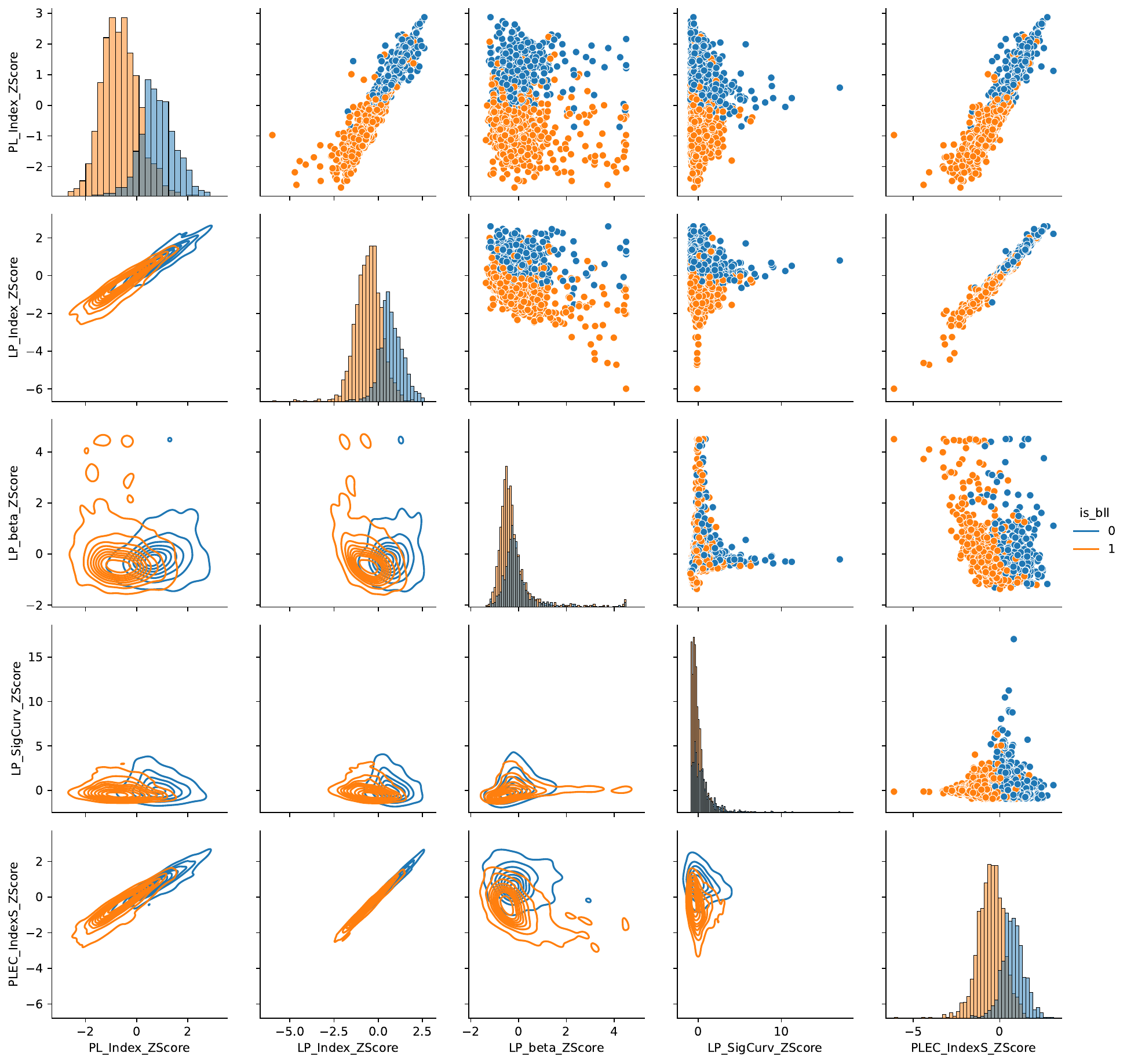}
\caption{Pair Grid Plot of Spectrum Variables, MICE.
The orange are the BLLs and the blue are the FSRQ.
}
\label{fig:sep1}
\end{figure*}

\begin{figure*}
\centering
\includegraphics[width=2.0\columnwidth]{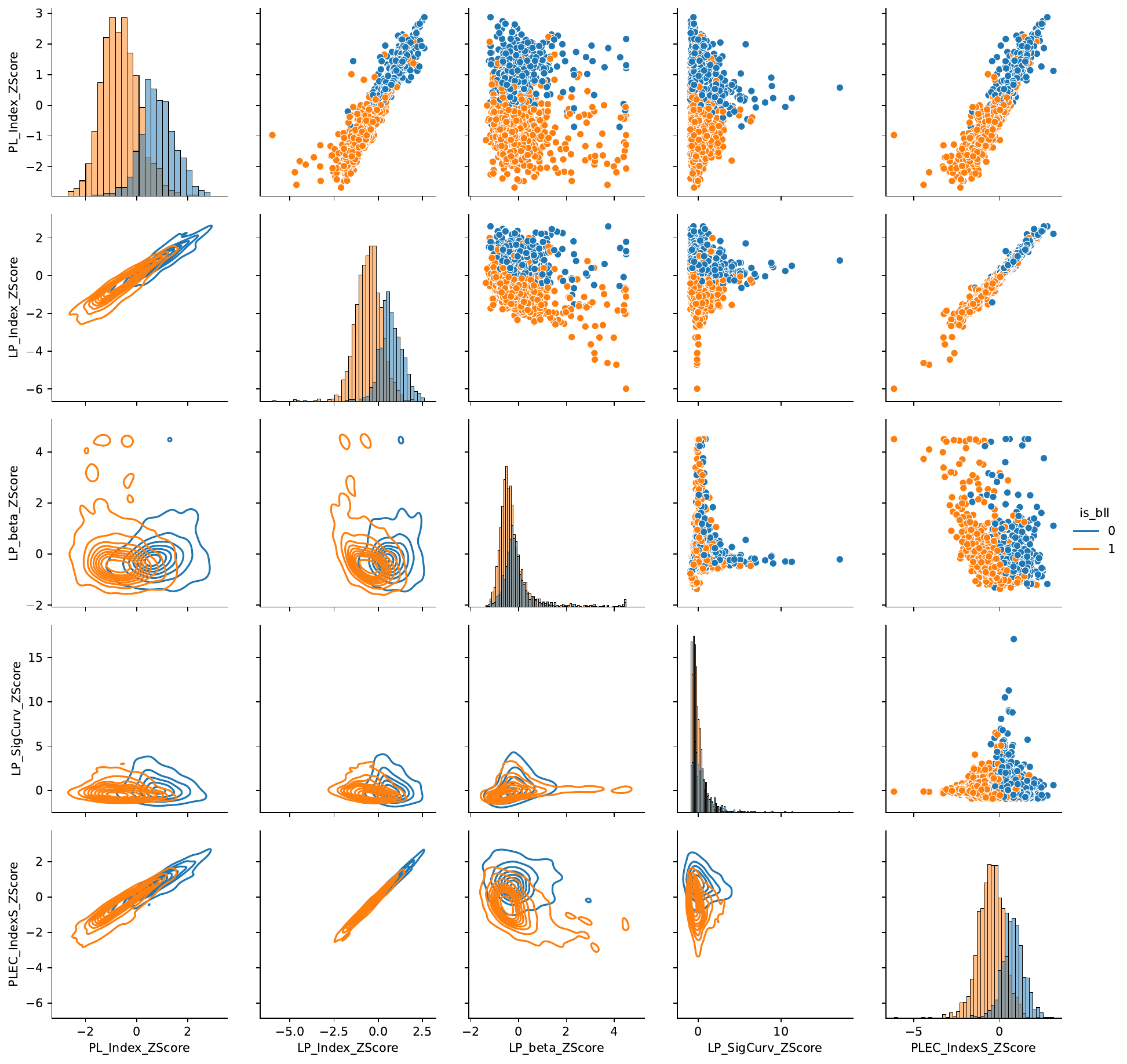}
\caption{Pair Grid Plot of Spectrum Variables, kNN. {\bf The orange are the BLLs, while blue the are FSRQs}}.
\label{fig:sep2}
\end{figure*}

Fig. \ref{fig:sep1} and \ref{fig:sep2} show a Pair Grid Plot. %produced using the Seaborn and Pandas libraries in Python. 
Along the diagonal are histograms of the z-score normalized data. Above the diagonal are scatter plots. 
Below the diagonal are kernel density plots, the contours showing the percentile of data within its boundary.
All data are color-coded such that BLLs are orange and FSRQs are blue. 
%The subset of data is presented in Fig. \ref{fig:sep1} and \ref{fig:sep2}.

\subsection{Machine Learning Models}

Upon imputing the missing data using kNN and MICE, we built several classification models to be used in the SuperLearner ensemble algorithm. 
As with handling the missing data, we used back-elimination to eliminate variables that had no statistically significant effect on the outcome of the logistic regression models. 
After logistic regression, the models we used are support vector Machines, Random Forests, Ranger Random Forests, neural networks, EARTH Regression, Bayes Regression, and extreme gradient boosting. We tested each algorithm separately for both KNN and MICE imputed data. 
Ultimately, SVMs, neural networks, and Bayes Regression were eliminated from the final MICE-imputed and kNN-imputed SuperLearner models. 
Below is a short description of the models included in the final MICE and kNN ensemble SuperLearner models. We use the default parameters of the R functions unless otherwise noted; in Bayes, Ranger, and Earth regressions we had good performance using the defaults. Our concern is that over-tuning of the parameters will bias the models to the training sets, and hinder performance when the models are eventually used against new data in future data releases.

\subsubsection{Parametric Methods: EARTH}
Enhanced Adaptive Regression Through Hinges (EARTH) is based on the Multivariate Adaptive Regression Splines method (MARS) \citep{FriedmanMARS}. The EARTH algorithm allows for better modeling of predictor interaction and non-linearity in the data compared to the linear model. 
EARTH fits a sum or product of hinges, which are part-wise linear fits of the data that are combined so that the sum-of-squares residual error is minimized with each added term.
In this study we set the family variable, i.e., the kernel to binomial.

\subsubsection{Decision Tree Methods: Random Forest, Extreme Gradient Boosting (XGB), and Ranger}

Random Forests are a supervised learning algorithm of the decision tree class. 
Decision trees measure conditional probabilities of the predictor variables with respect to the response variable and then sort the response variables according to the values of the individual predictor variables. 
This algorithm tends to over-fit the predictor variables \citep{randomForestsHo}. 
Random Forests are a method of compensating for the over-fitting of individual decision trees by randomly selecting subsets of data from the training data and creating many decision trees. The final prediction for the response variable is based on the majority outcome of the decision trees for a classification problem. 
In this study we set the max trees to 500 and the maximum number of nodes per tree to 10.

The Ranger algorithm is nearly the same as a standard Random Forest, except that it uses Extremely Randomized Trees (ERT). ERTs share the top-down procedure as Random Forests, except that the data cuts are randomly generated, assuming a uniform distribution of the predictor variables in the training set. 
The cut chosen is the one that minimizes prediction error \citep{geurts06extremetrees}.
The ranger function in R requires that the classification variable be set to true. 

Extreme Gradient Boosting (XGB) is also a decision tree class algorithm. 
We use the XGBoost library for R. Gradient Boosting creates an ensemble of decision trees as weak learners.  
These weak learners are then built into strong learners using an iterative gradient descent method to minimize a loss function such as root mean squared error (RMSE). 
XGB differs from other Gradient Boosting algorithms by using the Newton$-$Ralphson method instead of a gradient decent method. 
The Newton$-$Ralphson method is a root$-$finding algorithm used to minimize the loss function.
In this study, we set booster to "gbtree", eta to 0.001, max depth to 5, gamma to 3, subsample to 0.75, "colsample\_bytree" to 1, objective to "multi:softprob", "eval\_metric"  to "mlogloss" and "num\_class" to 2.
%Unlike other algorithms in our study, XGBoost does not use an R dataframe as the data input. It must be transformed into a Matrix and the algorithm requires a large number of parameters that are not set to a default value. 
%We use the parameters of "gbtree" as our booster, an "eta" of 0.001, "max depth" of 5, "gamma" of 3, a "subsample" of 0.75, "colsample{\_}by{\_}tree" of 1, and "objective" of "multi:softprob", "eval{\_}metric" of "mlogloss", and "num{\_}class" set to the number of classification levels, 2 for "bll" and "FSRQ" \citep{chen2016xgboost}. 

\subsubsection{Neural Network} %Apadted to more original language 
In Machine Learning, Artificial Neural Networks (ANN) are models that mimic biological neural networks. 
%Artificial neural networks (ANNs) for Machine Learning are a computational abstraction of biological neural networks. 
ANNs are a network similar to those in the mathematical discipline of Graph Theory in that the network is composed of nodes and edges. Nodes are the objects in the network, and edges are the connections between the objects. As ANNs are trained, statistical weights are assigned to edges between nodes that, based on input data, determine which pathway an observation takes through the network. These pathways will then determine the outcome. For an ANN classifier, the outcome is the predicted class of the observation \citep{jain1996}.
In this study, we set hidden node layers to 3 and 1, the activation function to logistic, and "linear.output" to false. %Biological neural networks train based on the rates at which signals go across synapses to neurons in the adjacent layer, with greater signal rates strengthening connections and lesser signal rates weakening connections. External stimuli affect these signal rates such that a specific set of stimuli will travel through the biological neural network and the creature's response to that stimuli depends on the pathways taken by the signals through the network \citep{jain1996}.} 

\subsubsection{Support Vector Machine} %[Wikipedia place holder deleted, adapted to more original language 

Support Vector Machines (SVMs) are a type of supervised machine-learning model that can be used for either classification or regression. %\sout{and are a supervised method} 
SVMs can classify multiple classes because they find boundaries between classes within the N-dimensional space created by the data, where N is the number of features in the data. 
The SVMs form these boundaries, called support vectors, using kernel functions such are linear kernels, polynomial kernels, Radial Basis Function (RBF) kernels, or sigmoid kernels. In this study, we used the SVM function from the e1071 library type was set to "C-classification", the radial kernel was used, the cost was set to 10, and the scale to false.  \citep{cortes1995support}.

\subsubsection{Bayes Regression} %[from Wikipedia - commented out adding more original language]

Bayes Regression is a conditional model that uses Bayes Theorem to predict a feature, such as the numerical class of an AGN, based on a linear combination of the other features of the data set. 
The regression coefficients are calculated via the posterior probability distribution of the features.
In this study, we use the normal linear model, which assumes the dependent feature is a function of a Gaussian distribution of the independent features \citep{raftery1996}.
The Gaussian distribution is well-suited for our study since we use z-score normalization, which measures the standard deviation from the mean for each feature.
%Since we use z-score normalization, which measures the standard deviation from the mean for each feature, the Gaussian distribution is well-suited for our study.

%Bayesian linear regression is a type of conditional modeling in which the mean of one variable is described by a linear combination of other variables, with the goal of obtaining the posterior probability of the regression coefficients (as well as other parameters describing the distribution of the regressand) and ultimately allowing the out-of-sample prediction of the regressand (often labeled 
%y) conditional on observed values of the regressors (usually 
%X). The simplest and most widely used version of this model is the normal linear model, in which 
%y given 
%X is distributed Gaussian. In this model, and under a particular choice of prior probabilities for the parameters—so-called conjugate priors—the posterior can be found analytically. With more arbitrarily chosen priors, the posteriors generally have to be approximated.

\subsubsection{SuperLearner}\label{sec:SL}

SuperLearner \citep{van2007super,Narendra2022,Dainotti2021,Gibson2022} is an ensemble method {that leverage on the advantages of several machine learning models and weights each constituent model according to its predictive power.}
It constructs models using a 10-fold cross-validation (CV). SuperLearner also calculates a set of normalized coefficients. 
The user defines the models used for SuperLearner's input and determines the final prediction by minimizing root-mean-squared error (RMSE) \citep{polley2010super}.  
Note that the final model is trained on 9 folds of the CV, and the $10^{th}$ fold is used on the test set, and this CV is internal to the SuperLearner function. The CV we perform in individual model testing is a separate procedure. The CV Normalized coefficients are higher for models with lower RMSE. 
The condition required by the SuperLearner is that the sum of the coefficients ($A_i$), which corresponds to the predictive power of the constituent models, should be 1.
We use these coefficients to determine which models will be included in the final model as explained below in section \ref{sec:MBP}, and the family variable was set to binomial. 

\subsection{Model Building Procedure}\label{sec:MBP}

The process for each algorithm is the same. The data have been separated into randomly selected training sets and test sets. We sorted 80\% of the data into the training set and 20\% into the test set. 
The models are trained on the training set. 
Randomly selected training/test sets are useful to prevent over-fitting the model.

%{Accuracy in ML classification is defined as the ratio of the number of correct predictions to the total number of data.}

When over$-$fitting occurs, the model exhibits high accuracy in classifying the training set but much lower accuracy when the model is used on new data. 
To address this issue, we built each model, regardless of the algorithm, using CV, see section \ref{sec:SL}. 
In CV, multiple training$-$test sets are created to test if the model is over$-$fitting to the training data and to quantify the effect of random selection between the training and test set on the final model. 
Then it is used to classify the objects in the test set.
If a model is not over$-$fitted, it should have high accuracy when classifying the test set.   
We use 100 nested 5-fold CV.
For each of the 100 iterations, the training/test sets are selected randomly, and the model's accuracy is measured on the test set. The histogram for 100 accuracy measurements is provided in Fig. \ref{fig:acc_hist}.

\begin{figure*}
\centering
\includegraphics[width=2.1\columnwidth]{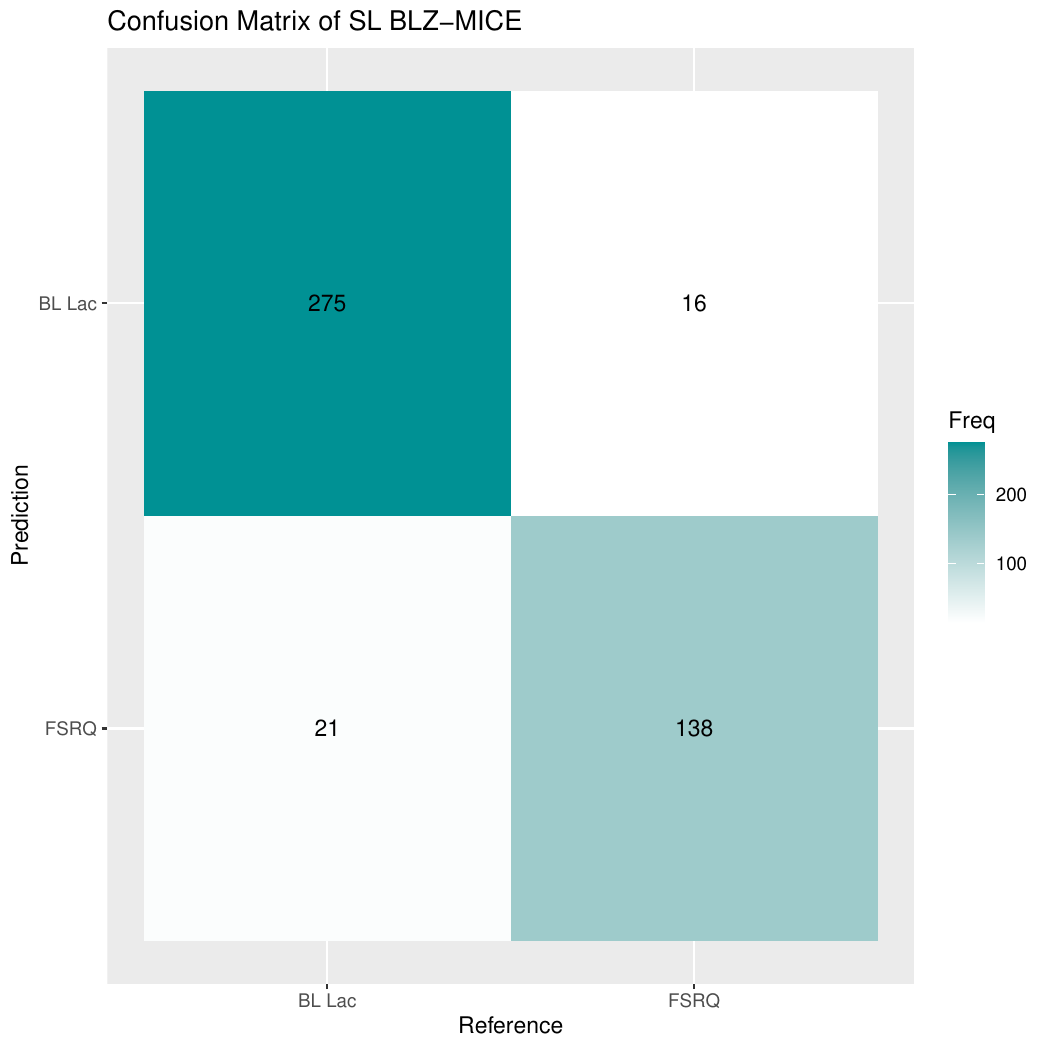}
\caption{Confusion Matrix Plot of an individual SuperLearner model with an accuracy of 91.8\%, a Sensitivity of 89.6\%, a Specificity of 92.9\%, and a Balanced Accuracy of 91.3\%. Reference indicates the class of the object in the 4FGL DR3 catalog. Prediction is the predicted class of the SuperLearner model.  
}
\label{fig:cm1}
\end{figure*}

We also created the confusion matrices \citep{Altman1552} for the performance of each model when validating the model using the test set. 
A confusion matrix is an $n \times n$ matrix, where $n$ is the number of classes that maps the actual classes on the column space and the predicted classes on the row space. The diagonal of the matrix are true predictions, i.e., BLL predicted to be BLL, and FSRQ predicted as FSRQs.
In Fig. \ref{fig:cm1} BLLs are mapped as 1 by the 'is{\_}bll' variable, the upper right are BLLs falsely identified as FSRQs, and the bottom left is for FSRQs falsely identified as BLLs.
The confusion matrix is useful for identifying if a model has a tendency for false positives (FP), FSRQs (is{\_}bll = 0) identified as BLLs ('is{\_}bll' = 1), or false negatives (FN), BLLs identified as FSRQs. 
From these confusion matrices, we calculate accuracy, which is the ratio of correct classifications from the test set to the size of the test set.

Similar to our procedure for logistic regression, we used the back$-$elimination process for each algorithm used in SuperLearner to select the optimal model for the final SuperLearner ensemble.
Back$-$elimination was based on the normalized coefficient that SuperLearner put on each algorithm, with those $< 5\%$ on average being eliminated. 
The following algorithms were not eliminated from the final model: EARTH Regression, XGB, Random Forests, and Ranger Random Forests. %and Neural Nets (kNN only). 
Both MICE and kNN imputed SuperLearner Models used stepAIC for predictor variable selection, which is an independent process from the model selection. 

StepAIC works similarly to back$-$elimination in that non-significant predictor variables are removed from the model. However, the statistic used is the Akaike information criterion (AIC) \citep{sakamoto1986akaike}:

\begin{equation}
AIC = 2(k -ln(\hat{L})
\end{equation}

where k is the number of parameters in the model and $\hat{L}$ is the maximized likelihood function; the probability function that the parameters yield the correct result \citep{van2007super}.  %SuperLearner reference. 

\section{Results}\label{sec:results}

%Add Pics
%First significant results
%Then non-significant 

SuperLearner performed well with both MICE and kNN imputation of missing data, with 91.2\% and 91.1\% average accuracy, respectively, on the test set. 
As mentioned in section \ref{sec:handling}, there is an IR of 1.84.
%times as many BLLs to FSRQs in the training and test set, this is called the Imbalance Ratio (IR). 
The SuperLearner model with MICE imputed data predicted that of the 1519 BCUs, 890 are BLLs, and 629 are FSRQs for an IR of 1.41. %BLLs to FSRQs. 
The SuperLearner model with kNN imputed data predicted 892 BLLs and 627 FSRQs for an IR of 1.42.% BLLs to FSRQs.  
%(see top right and bottom left panels of Fig. \ref{fig:class_bar}).
%Figs. \ref{fig:b1} and \ref{fig:c1}. 
Despite the imputation method, SuperLearner predicts a slightly lower ratio of BLLs in the unclassified blazars than in the classified set. The accuracies of the other models are shown in Table \ref{tab:model_acc}.

The logistic regression model using kNN imputed data had an accuracy of 90.3\%. The statistically significant variables were Fractional Variability, is{\_}PL, PL index, and Redshift.
The logistic regression model trained on kNN imputed data classified the BCUs into 900 BLLs and 619 FSRQs for an IR of 1.45.% BLLs to FSRQs.
%When the kNN logistic regression model was used to predict the classification of the BCUs, it resulted in 900 BLLs and 619 FSRQs for a ratio of 1.45 BLLs to FSRQs. 

The logistic regression model with MICE$-$imputed data has an accuracy of 90.3\%. The statistically significant variables are Fractional Variability, Gaia G Magnitude, PL index, LP SigCurv, and Redshift.
When the MICE logistic regression model is used to predict the classification of the BCUs, it results in 875 BLLs and 644 FSRQs for an IR of 1.36.% BLLs to FSRQs.

The logistic regression model with missing data removed by feature has an accuracy of 89.8\%.
The statistically significant variables are Variability Index, Frac Variability, is{\_}PL, and PL index. When the logistic regression model is used to predict the classification of BCUs, it results in 896 BLLS and 623 FSRQs for an IR of 1.44.% BLLs to FSRQs.  
The logistic regression model missing data removed by observation has an accuracy of 89.2\%. The IR was 1.38. There are only 12 BCUs with complete data, and the model predicts 10 FSRQs and 2 BLLs. Note that the weak spectra of BLLs biases this result.

As stated above, the data have an IR of 1.84. In dealing with imbalance data, \cite{lopez2013} note that problems with imbalanced data often do not arise from the IR itself but from other issues with the data, such as overlapping class separability, which we can see in Figs \ref{fig:sep1} and \ref{fig:sep2}. Their analysis for overlapping class separability starts with a data set with an IR of 9. In \cite{johnson2019}, a meta-analysis of studies on machine learning with high IR, the smallest ratio of IR in the meta-analysis is 2. Taken together, it seems that an IR of 1.84 is below the threshold for using special techniques other than using other statistics than simple accuracy, which we will discuss below.   

\begin{figure*}
%\centering
    \includegraphics[width=\columnwidth]{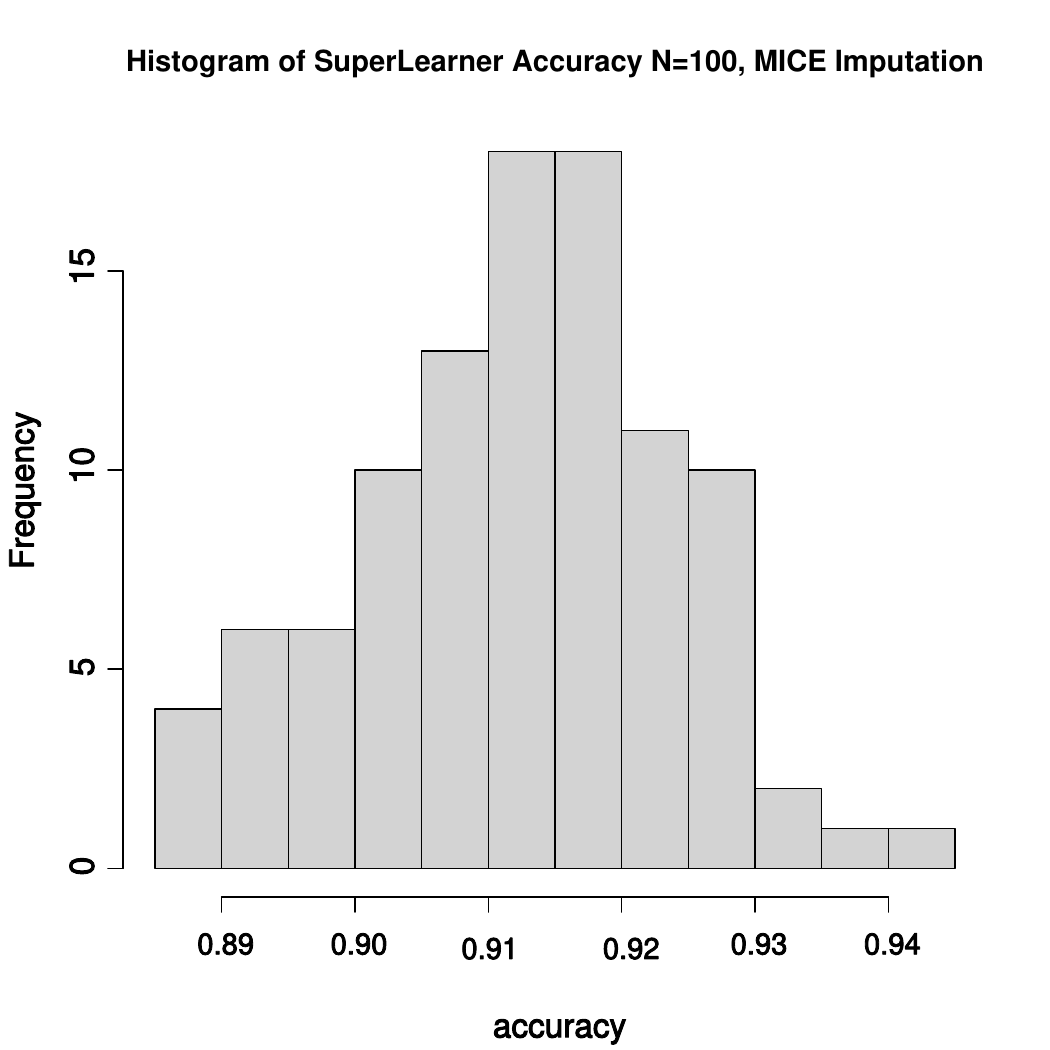}
    \includegraphics[width=\columnwidth]{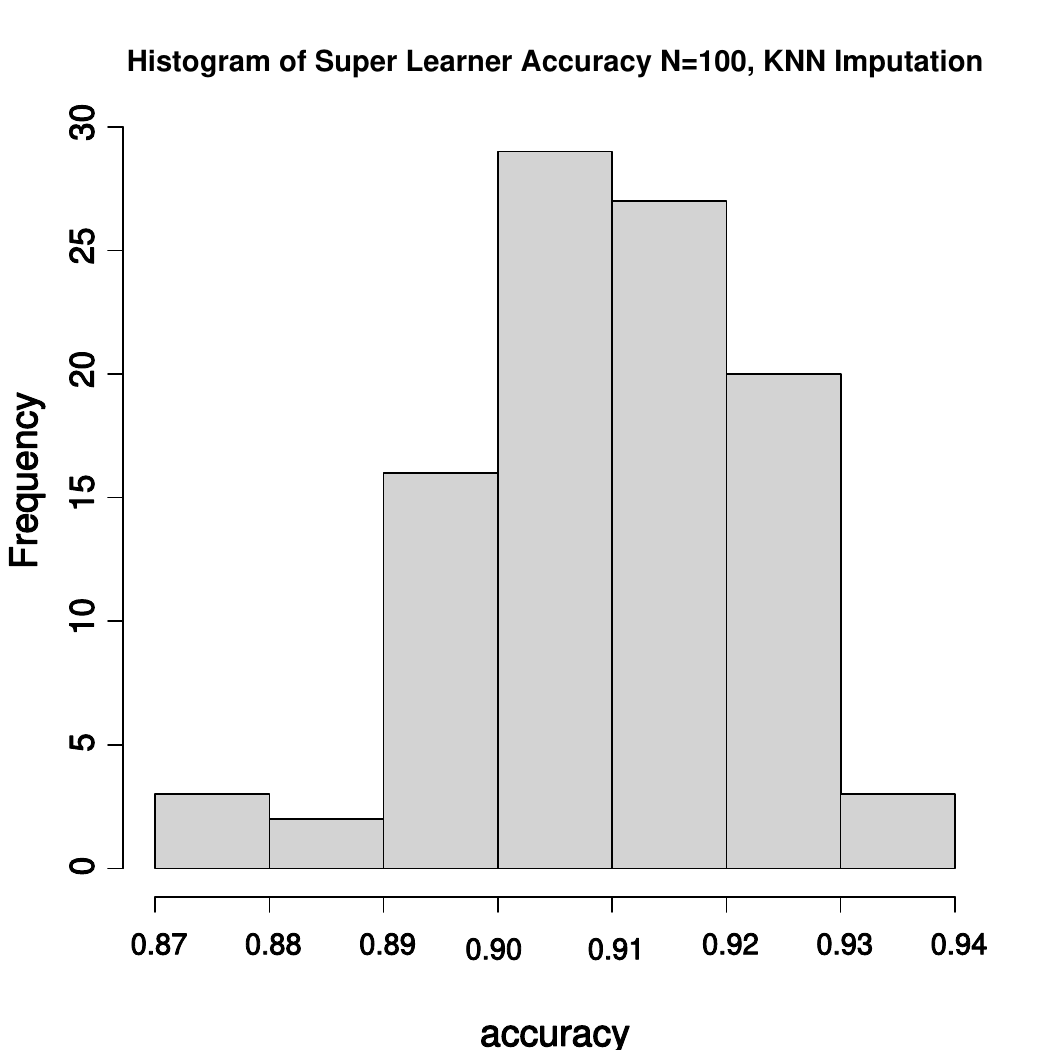}
    \includegraphics[width=\columnwidth]{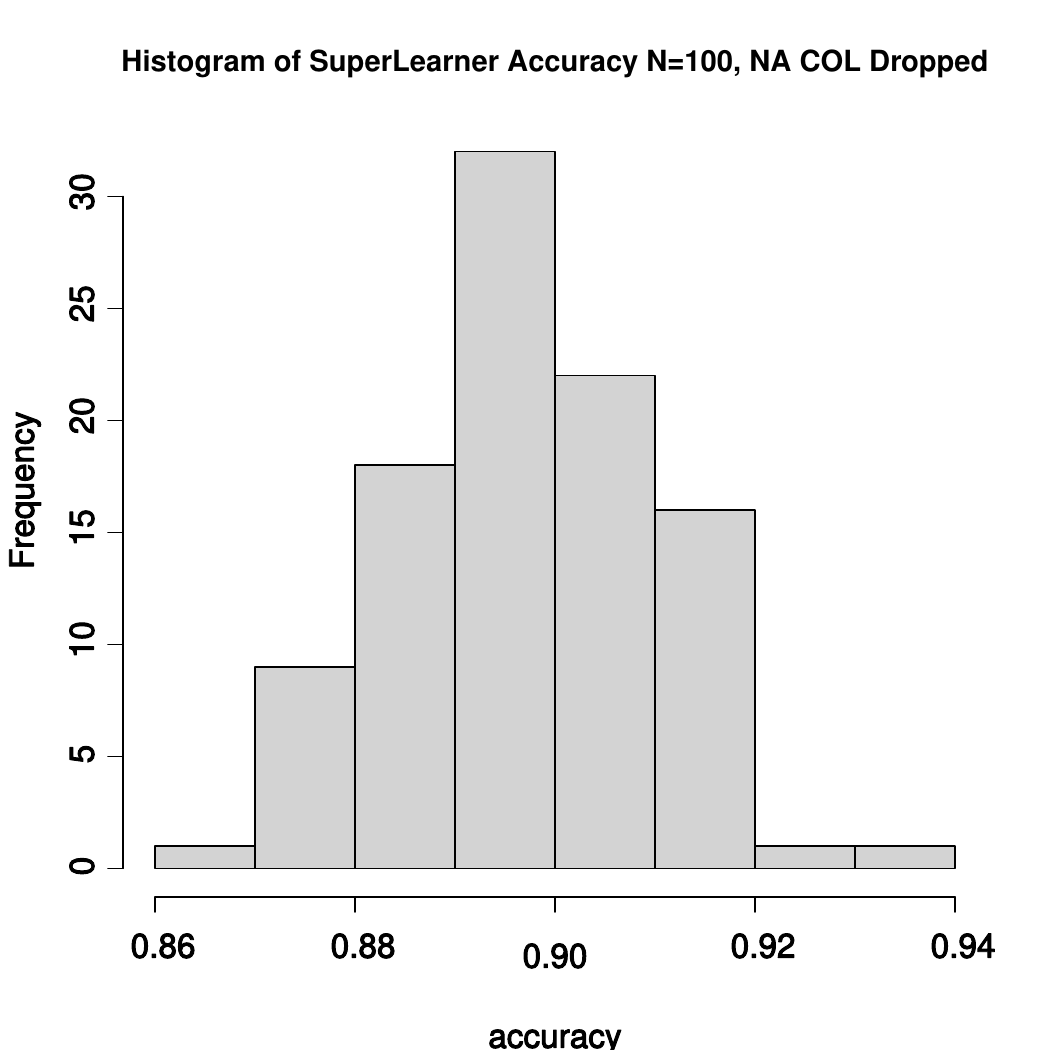}
    \centering
    \caption{Histograms of SuperLearner Model Accuracy.
    Top left: SuperLearner with MICE imputed data.
    Top right: SuperLearner with kNN imputed data.
    Bottom: SuperLearner with NA columns dropped.
    }
    \label{fig:acc_hist}
\end{figure*}

The accuracy of each model averaged over 100 times nested 5$-$fold CV, is shown in Table \ref{tab:model_acc}; the histograms of the SuperLearner models are shown in Fig. \ref{fig:acc_hist}. 
Please note that the MICE method tends to underperform kNN, albeit not beyond fluctuation due to the random selection of training and test sets. 
Imputing missing data outperforms excluding missing data in the logistic regression and SuperLearner models.  
From Table \ref{tab:model_acc}, we can see that the best-performing models are Ranger, XGB, and SuperLearner. 
Overall there is not a large difference among the models, although the SuperLearner, as expected, has one of the highest predictions.

We also have created a SuperLearner model using the data where columns with missing data are dropped, as these data are performed the same as in the logistic regression model. 
When the missing data are not accounted for, SuperLearner Model has an accuracy of 89.8\% under nested 5-fold CV, see bottom panel of Fig. \ref{fig:acc_hist}.
This is only slightly less accurate than the imputed data models, which are 91.2\% and 91.1\% for MICE and kNN, respectively. 
When used to predict the class of the BCUs, SuperLearner predicts 890 BLLs and 629 FSRQs for a ratio of 1.41. %(see bottom right panel of Fig. \ref{fig:class_bar}).
Regardless of the data treatment, SuperLearner predicts a lower ratio of BLLs to FSRQs than the data in the training and test sets.

Table \ref{tab:SL_stats} shows the classification statistics for the MICE Imputed SuperLearner Classifier averaged over the 100 nested CVs. 
The confusion matrix function in R interprets FSRQs as positive and BLLs as negative. 
Sensitivity is also called the True Positive Rate; this means the ratio of correctly classified FSRQs to total FSRQs in the test set is 87\%. 
Specificity is also called the True Negative Rate. This is the rate of correctly classified BLLs to total BLLs in the test set, 94\%.
Balanced Accuracy is the arithmetic average of these two scores, 90\%. 
Note that the accuracy score reported in Table \ref{tab:model_acc} is the number of correct classifications over the sample size. 
The two scores do not agree because BLLs outnumber FSRQs by 1457 to 794. Given this class imbalance, the balanced accuracy is the more reliable of the two accuracy measures.  
Positive Predictive Value (PPV) is the ratio of correctly identified FSRQs to the sum of correctly identified FSRQs and BLLs falsely identified as FSRSs, 89\%, and Negative Predictive Value is the ratio of correctly identified BLLs to the sum of correctly identified BLLs to FSRQs falsely identified as FSRQs, 93\%. The F1 Score is the harmonic mean of the PPV and sensitivity. The F1 score is calculated by:

\begin{equation}
    F_{1} = \frac{2*PPV*sensitivity}{PPV+sensitivity}
\end{equation}

In Statistics, harmonic means, such as the F1 score, are used to calculate the average rate, here F1 is 88\%.
%In this instance, the average between PPV and sensitivity is the average between correctly classified FSRQs to all predicted FSRQs, 89\%, and correctly classified FSRQs to all actual FSRQs in the sample, 87\%.  }

Prevalence is the number of FSRQs in the sample over the total sample size, 35\%. The detection rate is the number of correctly predicted FSRQs over the total sample size, 31\%. 
The detection prevalence is the number of all predicted FSRQs over the sample size, 35\%.

\begin{table}
    \centering
    \caption{Model Accuracy, 5$-$fold CV, 10$x$nested for Neural networks, 100$x$nested for other models}
    \label{tab:model_acc}
    \begin{tabular}{c|c|c}
    \hline
    \hline
    Model & MICE Accuracy & kNN Accuracy \\
    \hline
     Logistic regression & 90.3\% & 90.3\% \\
     Random Forest & 91.1\% & 90.8\% \\
     SVM & 90.0\% & 90.3\% \\
     XGB & 91.2\% & 91.4\%   \\
     Bayes Regression & 90.4\% & 90.3\% \\
     EARTH & 89.9\% & 90.2\% \\
     Ranger & 91.0\% & 91.1\%\\
     Neural networks & 89.7\% & 90.0\% \\
     SuperLearner & 91.2\% & 91.1\%\\
    \end{tabular}
\end{table}

\begin{table}%[ht]
    \centering
    \caption{Classifaction Statistics, 5$-$fold CV, 100$x$nested SuperLearner MICE Imputation}
    \label{tab:SL_stats}
    \begin{tabular}{rr}
    \hline
    Statistic & Score \\ 
    \hline
  Sensitivity & 0.87 \\ 
  Specificity & 0.94 \\ 
  Pos Pred Value & 0.89 \\ 
  Neg Pred Value & 0.93 \\  
  F1 & 0.88 \\ 
  Prevalence & 0.35 \\ 
  Detection Rate & 0.31 \\ 
  Detection Prevalence & 0.35 \\ 
  Balanced Accuracy & 0.90 \\ 
   \hline
   \end{tabular}
\end{table}

%\begin{figure}
%     \centering
%     \begin{subfigure}[b]{0.45\columnwidth}
%         \centering
%         \includegraphics[width=\columnwidth]{ACC_DR3_SL_MICE_Hist.pdf}
%         \caption{MICE}
%         \label{fig:a3}
%     \end{subfigure}
%     \hfill
%     \begin{subfigure}[b]{0.45\columnwidth}
%         \centering
%         \includegraphics[width=\columnwidth]{DR3_SL_kNN_acc_hist.pdf}
%         \caption{kNN}
%         \label{fig:b3}
%     \end{subfigure}
%      \begin{subfigure}[b]{0.45\columnwidth}
 %        \centering
  %       \includegraphics[width=\columnwidth]{ACC_DR3_SL_NAC_Hist.pdf}
%         \caption{NA Columns Dropped}
 %        \label{fig:c3}
  %   \end{subfigure}
   %     \caption{Histograms of SuperLearner Model Accuracy.}
    %    \label{fig:acc_hist2}
%\end{figure}

%\subsection{Classification with multiple AGN classes}
%ADD THE CONFUSION MATRIX FOR ALL CLASSES AND THE CONFUSION MATRIX FOR %THE THREE CLASSES AND THE EXPLANATION. 
\section{Summary and Conclusions}\label{sec:conclusions}
Our aim is to classify AGN using several classification techniques embedded in the SuperLearner, such as EARTH Regression, XGBoost, Random Forests, and Ranger Random Forests. To allow the maximum number of observations with all the features, we use MICE. We also use kNN imputation as a comparison to MICE.  We here summarize below our major findings, which are relevant both for the classification methodology and from a physical point of view: 
\begin{itemize}
    \item MICE Imputation is statistically tied with kNN imputation with every algorithm.

     \item Using imputation of missing data outperforms dropping data with missing variables either along features (columns) or observations (rows). 

     %• Gaia G Magnitude becoming significant when we excluded Redshift may be due to the Magnitude's dependence on the distance to the source, for which Redshift is a measure. \\

     \item Spectral parameters, such as PL Index and LP SigCurv, and variability parameters, such as Fractional Variability, are significant in logistic regression modeling.  
    
     \item 100 GeV $\gamma$-ray flux is not significant in any logistic regression model.

    \item Redshift’s significance in classification may support the notion that BLL and FSRQs, and by extension, FR1 and FR2 radio galaxies, may be evolutionary steps in radio-loud AGNs.
     
    \item Although XGB has slightly higher accuracy, note that this could be due to statistical fluctuation, we expect SuperLearner to perform better on new data, as ensemble methods are less likely to over-train on the training data. 
\end{itemize}

   Redshift’s significance in classification paves the way for a better understanding of this class of objects. In addition, we use the SuperLearner for the first time in the realm of classification.
   We point out that SuperLearner has already been successfully used in the realm of regression \citep{Narendra2022,Dainotti2021,Gibson2022}.
   The use of SuperLearner here opens a new perspective on astrostatistics analysis. It allows the application of powerful machine-learning methods that are still unknown or rarely used in the astronomy community. 
   When new data is available and the new AGN catalog is released, we will be able to further support this scenario or highlight new subtle features which otherwise would remain buried in the classification.

\section*{Acknowledgements}

This research was supported by the Visibility and Mobility module of the Jagiellonian University (Grant number: WSPR.WSDNSP.2.5.2022.5) and the NAWA STER Mobility Grant (Number: PPI/STE/2020/1/00029/U/00001). A. N. is grateful to the Exploratory Research Grant for the financial support for the visit of A. Narendra to the Division of Science at the National Astronomical Observatory of Japan.

%The Acknowledgements section is not numbered. Here you can thank helpful
%colleagues, acknowledge funding agencies, telescopes and facilities used etc.
%Try to keep it short.

%%%%%%%%%%%%%%%%%%%%%%%%%%%%%%%%%%%%%%%%%%%%%%%%%%
\section*{Data Availability}

The data used in this paper is taken from the Fourth LAT AGN Catalog \citep{2022ApJS..260...53A}. Post-processed data is available upon request.

%%%%%%%%%%%%%%%%%%%% REFERENCES %%%%%%%%%%%%%%%%%%

% The best way to enter references is to use BibTeX:

\bibliographystyle{mnras}
%\bibliography{refs} % if your bibtex file is called example.bib

% Alternatively you could enter them by hand, like this:
% This method is tedious and prone to error if you have lots of references
%\begin{thebibliography}{99}
%\bibitem[\protect\citeauthoryear{Author}{2012}]{Author2012}
%Author A.~N., 2013, Journal of Improbable Astronomy, 1, 1
%\bibitem[\protect\citeauthoryear{Others}{2013}]{Others2013}
%Others S., 2012, Journal of Interesting Stuff, 17, 198
%\end{thebibliography}

%%%%%%%%%%%%%%%%%%%%%%%%%%%%%%%%%%%%%%%%%%%%%%%%%%

%%%%%%%%%%%%%%%%% APPENDICES %%%%%%%%%%%%%%%%%%%%%

%\appendix

%\section{Some extra material}

%If you want to present additional material which would interrupt the flow of the main paper,
%it can be placed in an Appendix which appears after the list of references.

%%%%%%%%%%%%%%%%%%%%%%%%%%%%%%%%%%%%%%%%%%%%%%%%%%

\section{Appendix: Classification of All AGN Types}\label{sec:appdx}

This Appendix shows SuperLearner's performance when all identified AGN types are included. The AGN classes included in DR3 are BLL, FSRQs, BCUs, as above, as well as 45 Radio Galaxies (RDG), 9 Active Galactic Nuclei (AGN) with no further sub-type, 8 Narrow$-$line Seyfert 1's (NLSY1), 5 Compact Steep Spectrum (CSS), 2 Seyferts (SEY), and 2 Steep Spectrum Radio Quasars (SSRQs).

For logistic regression classifiers, the accepted practice is to have ten observations for one feature \citep{peduzzi1996}.
We use eighteen features in this analysis. Under the accepted practice, the minimum number of observations needed for a reliable result is 180. Only BLLs and FSRQs meet that threshold, with BCUs being withheld for classification. 
However, the accepted 10:1 ratio was established using solely logistic regression. SuperLearner is an ensemble method. Below we test if SuperLearner can perform well with fewer than 10 observations per feature.

Of the non$-$blazar classes, only the 45 RDGs have a count high enough to be represented in both the training and test sets. Under binomial probability distribution, the expected number of RDGs in the training set is 36, and there is a 0.004\% that no RDGs appear in the test set. 
However, for the 9 AGN, the expected number of AGN in the training set is 7.2, and the probability of no AGN in the test set is 13.4\%.

SuperLearner requires that the training and test sets have the same number of classes. SuperLearner also only supports binomial classification, that is, classification between two classes. 
To meet these two requirements, we created three binary classifiers. As with the blazar analysis, we created "is\_bll" with 1 representing a BLL and 0 representing another class. 
Additionally, we created "is\_fsrq" where 1 is for FSRQs and 0 for non-FSRQs, and finally, we created "is\_rdg" where 1 is for RDGs and 0 for another class. The other classes, AGNs, NLSY1s, CSSs, SEYs, and SSRQs, with too few observations to reliably appear in both the training and test sets, were reclassified as "NBAGN" for non-blazar AGN.

During the cross$-$validation loop, we trained three instances of SuperLearner, one to classify BLLs, one to classify FSRQs, and one to classify RDGs. Each class is represented by a binary vector: (1,0,0) for BLLs, (0,1,0) for FSRQs, (0,0,1) for RDGs, and all other combinations are classified as NBAGNS. 
The resultant classification statistics are in Table \ref{tab:SL_agn_stats}. Note with small numbers of RDGs and NBAGNs, if low numbers of these objects were contained in the test sets, PPV, NPV, and Prevalence and, by extension, F1 value can have division by zero, resulting in a not a number (NaN) value. 
For NBAGNS, due to the low count, sensitivity or PPV for an individual instance in the CV can be zero, resulting in an F1 value of NaN since F1 is proportional to $\frac{1}{PPV + sensitivity}$. 
For RDGs, the PPV is NaN resulting in an F1 value of NaN.%, as F1 also goes as $\frac{1}{PPV}$. }

We see in Table \ref{tab:SL_agn_stats} that the Sensitivity for RDGs is 6\%, and the Sensitivity for NBAGNs is 9\%. 
Specificity is 97\% for NBAGNs and 100\% upto two significant figures for RDGs. 
So, while the model does not accurately predict RDGs or NBAGNs, they tend not to misclassify other objects as RDGs or NBAGNs. 
Note that the performance of the models for BLLs and FSRQs is comparable to the binomial classifier, see Table \ref{tab:SL_stats}. 
Specifically, since the binomial classifier treated FSRQs positive cases comparing Table \ref{tab:SL_stats} to the FSRQ row in Table \ref{tab:SL_agn_stats}, we see that FSRQs' multinomial Sensitivity is 87\%, Specificity is 94\%, PPV is 87\%, NPV is 92\%, F1 is 86\%, Prevalence is 34\%, Detection Rate is 29\%, Detection Prevalence is 33\% and Balanced Accuracy is 88\%.
These stats are all approximately 1 to 2\% less than the Binomial case, and the combined number of RDGs and NBAGNs represents 3.1\% of the total sample. This suggests that the RDGs and NBAGNs misclassified as Blazar types bring these average statistics down. 

\begin{figure*}
\centering
\includegraphics[width=2.1\columnwidth]{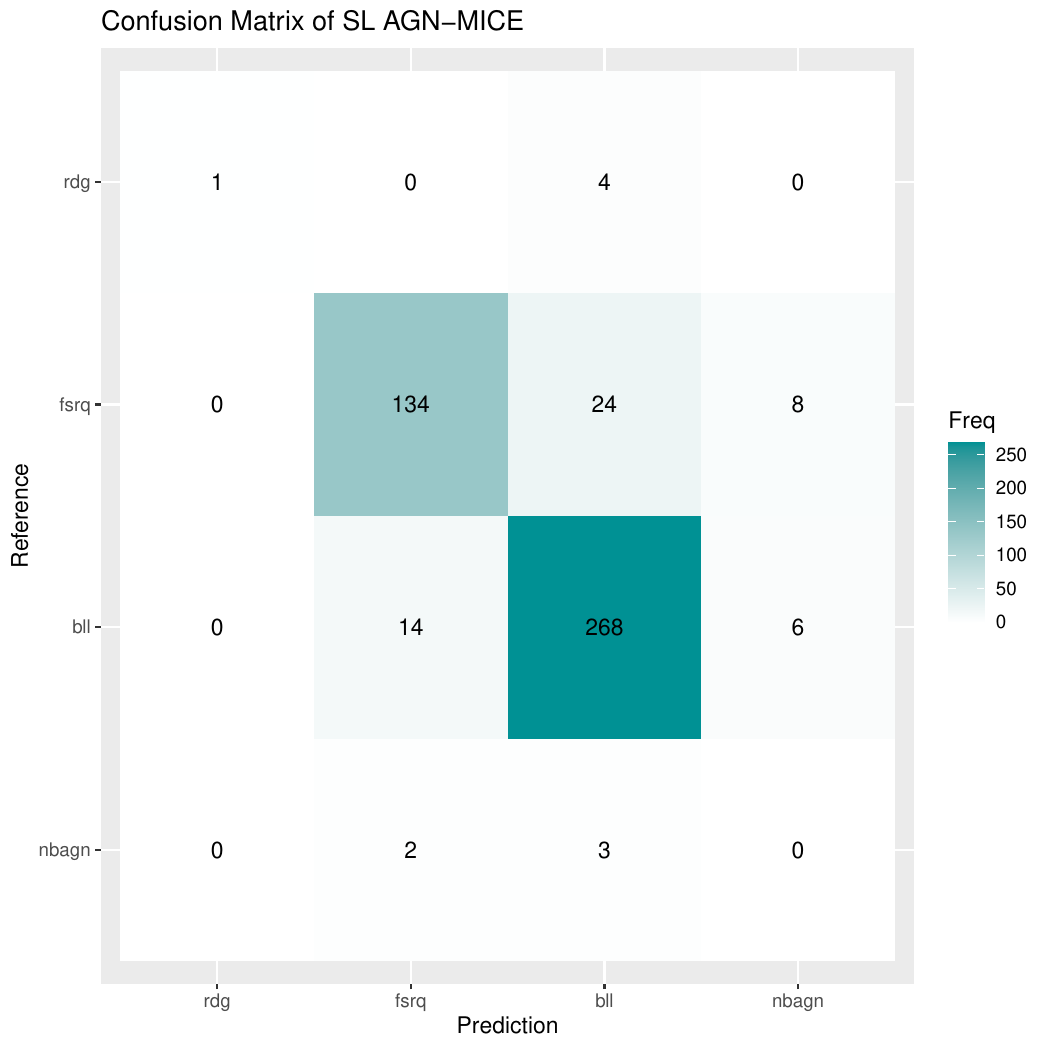}
\caption{Confusion Matrix Plot of an individual SuperLearner model with an accuracy of 86.9\%, a Sensitivity of 89.6\%, a Specificity of 92.9\%, and a Balanced Accuracy of 91.3\%. The reference indicates the class of the object in the 4FGL DR3 catalog. Prediction is the predicted class of the SuperLearner model.  
}
\label{fig:cm2}
\end{figure*}

Figure \ref{fig:cm2} shows the Confusion Matrix corresponding to one model. 
In this model, NBAGN Sensitivity is 0\% with 0 True Positives and 5 False Negatives, and Specificity is 97\% with 403 True Negatives and 14 False Positives; 
RDG Sensitivity is 20\% with 1 True Positive and 4 False Negatives, and Specificity is 100\% with 402 True Negatives and 0 False Positives; 
BLLs have a Sensitivity of 93\% with 268 True Positives and twenty False Negatives, and a Specifity of 81\% with 135 True Negatives and 31 False Positives; 
FSRQs have a Sensitivity of 81\% with 134 True Positives and 32 False Negatives and a Specificity of 94\% with 269 True Negatives and 16 False Positives. 
It is also noteworthy how this uneven class count affects Sensitivity and Specificity with the more numerous class, BLLs being weighted toward higher Sensitivity and lower Specificity, and FSRQs, RDGs, and NBAGN being weighted toward higher Specificity and lower Sensitivity. 
It is also worth noting that this model demonstrates the 100\% Specificity of the RDGs, which means the model does not misidentify other classes as RDGs.

We conclude in this Appendix that these data support the findings of \cite{peduzzi1996} regarding observation to features ratio.
To accurately and positively predict other AGN categories, more observations of RDGs and the types that make NBAGNs are needed. Although the 45 observations of RDGs do seem to be enough to prevent False Positives of that class.

\begin{table*}
    \centering
    \caption{Classifaction Statistics, 5$-$fold CV, 100$x$nested SuperLearner MICE Imputation - All AGN}
    \label{tab:SL_agn_stats}
    \begin{tabular}{llllllllll}
     \hline
     & Sensitivity & Specificity & PPV & NPV & F1 & Prevalence & Detection Rate & Detection Prevalence & Balanced Accuracy \\ 
     \hline
    Class: nbagn & 0.09 & 0.97 & 0.04 & 0.99 & NaN & 0.01 & 0.00 & 0.03 & 0.53 \\ 
    Class: bll & 0.93 & 0.84 & 0.91 & 0.87 & 0.92 & 0.63 & 0.58 & 0.64 & 0.88 \\ 
    Class: fsrq & 0.85 & 0.94 & 0.87 & 0.92 & 0.86 & 0.34 & 0.29 & 0.33 & 0.89 \\ 
    Class: rdg & 0.06 & 1.00 & NaN & 0.98 & NaN & 0.02 & 0.00 & 0.00 & 0.53 \\ 
    \hline
    \end{tabular}
\end{table*}

% Don't change these lines
\bsp	% typesetting comment
\label{lastpage}
\end{document}